\documentclass[aps,groupedaddress,twocolumn,pra]{revtex4-2}
\usepackage{soul}
\usepackage{dsfont}
\usepackage[utf8]{inputenc}
\usepackage{physics}
\usepackage{braket}
\usepackage{amsmath}
\usepackage{amssymb}
\usepackage{dsfont}
\usepackage{appendix}
\usepackage{xcolor}
\usepackage{graphicx,bbm}
\usepackage{comment}
\usepackage[hidelinks]{hyperref}
\hypersetup{
	colorlinks,
	linkcolor={red!50!black},
	citecolor={blue!50!black},
	urlcolor={blue!80!black}
}
\usepackage{mathtools}
\newcommand{\ep}[1]{\text{\tiny{EP}}}

\newcommand{\braa}[1]{\langle\!\langle #1 |}
\newcommand{\brakett}[2]{\langle\!\langle #1 | #2 \rangle\!\rangle}
\newcommand{\ketbraa}[2]{| #1 \rangle\!\rangle \langle\!\langle #2 |}
\newcommand{\pp}{\text{\tiny{++}}}
\newcommand{\mm}{\text{\scriptsize{-}\hspace{0.02cm}\scriptsize{-}}}

\newcommand{\dket}[1]{|#1\rangle \! \rangle }
\newcommand{\dbra}[1]{\langle \! \langle #1| }

\newcommand{\dketbra}[2]{|#1\rangle \! \rangle \langle \! \langle #2| }
\newcommand{\dbraket}[2]{ \langle \! \langle #1|#2\rangle \! \rangle }

\newcommand{\kett}[1]{| #1 \rangle\!\rangle}

\begin{document}
\title{Framework for (non-)adiabatic chiral state conversion: from non-Hermitian Hamiltonians to Liouvillians}
\author{Elna Svegborn}

	\affiliation{Physics Department and NanoLund, Lund University, Box 118, 22100 Lund, Sweden}
\author{Shishir Khandelwal}
	\affiliation{Physics Department and NanoLund, Lund University, Box 118, 22100 Lund, Sweden}

\begin{abstract}
Adiabatic chiral state conversion (CSC) is one of the many counterintuitive effects associated with non-Hermitian physics. In quantum systems, numerous works have demonstrated this phenomenon under both non-Hermitian Hamiltonian and Lindblad evolution. However, despite considerable progress, the physical mechanism behind it has been a subject of debate. In this work, we present a unified framework that explains CSC in any non-Hermitian system, encompassing non-Hermitian Hamiltonian, Lindblad, and hybrid settings. Our framework relies on perturbative, non-adiabatic corrections to adiabatic evolution and consistently predicts CSC with only the lowest-order corrections. We demonstrate its efficacy with models of single and coupled dissipative qubits, obtaining analytical solutions for the conversion fidelity. Our analysis further reveals the role of non-perturbative dynamics, which can be present even in apparently slow trajectories. We show that this property can be utilised to considerably enhance state conversion. Finally, we demonstrate that CSC can be observed in a model without the presence of exceptional points.    
\end{abstract}

% \date{\today}
\maketitle

\section{Introduction}

Adiabatic evolution in Hermitian and non-Hermitian systems is fundamentally different \cite{Garrison1988}. While adiabatic variation of parameters in Hermitian Hamiltonians results in evolution along instantaneous eigenstates, the same process in non-Hermitian Hamiltonians (NHHs) may induce switching between eigenstates \cite{Dembowski2001,Stehmann2004}. For closed parameter trajectories in the non-Hermitian case, the resulting eigenstate switching has been shown to exhibit a chirality \cite{Uzdin2011,Gilary2013}, giving rise to the phenomenon known as \textit{chiral state conversion} (CSC). Initially, this effect was attributed to encircling non-Hermitian degeneracies called exceptional points (EPs) \cite{Uzdin2011,Berry2011}, with experiments confirming this observation \cite{Doppler2016}. However, it was soon understood that similar observations can be made in the vicinity of an EP, while not necessarily encircling it \cite{Nasari2022,Hassan2017}. Although the exact mechanism behind CSC has been a subject of debate, it is known that the effect strongly depends on the degree of adiabaticity, underscoring the importance of non-adiabatic considerations \cite{Wang_2018,Hassan2017,Nye2023,Nye2024}. Recently, Ref. \cite{Kumar2025} has given a description of CSC under NHH evolution through non-adiabatic corrections along with uncontrolled fluctuations.

A natural platform for investigating the physics of chiral state conversion is open quantum systems, as their evolution is inherently non-Hermitian. Here, it is necessary to go beyond the NHH, since it captures only a part of the full picture, and consider full Lindblad dynamics \cite{Minganti2019,Khandelwal2021}. Accordingly, CSC has been predicted \cite{Kumar2021,Sun2023,Sun2024,Khandelwal2024a,Wu2025} and observed \cite{Chen2022,Chen2025,Bu2023,Bu2024,Gao2025,Tang2025} under Lindblad dynamics. However, a consistent theory that gives both a qualitative understanding, as well as quantitative predictions of the effect in these settings is missing. In this work, we aim to bridge this gap by developing a framework for state conversion in arbitrary non-Hermitian systems. To this end, we introduce a unifying framework for slow driving in open quantum systems, which naturally encompasses NHH, Lindblad and hybrid-Lindblad \cite{Minganti2020} evolution. We then apply the framework to study CSC in single and coupled dissipative qubits, revealing the origin of CSC and the physical conditions necessary for it. In the Lindblad case, we further obtain analytical expressions for state conversion fidelities and discuss the possible benefits of non-perturbative dynamics for enhancing the effect. Finally, we present a model that exhibits CSC despite having no EPs; this goes against the conventional understanding that CSC requires parameter trajectories in the vicinity of an EP. 

\begin{figure}
    \centering
    \includegraphics[width=0.75\columnwidth]{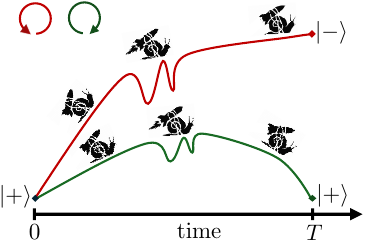}
    \caption{A sketch of chiral state conversion. A non-Hermitian system is slowly driven from one eigenstate; depending on the orientation of the parameter trajectory, it either makes a non-adiabatic transition to another eigenstate or returns to the initial one. Even though the evolution may involve noisy or far-from-adiabatic behaviour along the trajectory due to its form or the inherent properties of the involved dynamics, the conversion is well described by slow-driving predictions.  }
    \label{fig:placeholder}
\end{figure}

We begin in Sec. \ref{sec:prelim} by giving preliminaries regarding NHH and Lindblad dynamics, as well as the models that we investigate in this work. In Sec. \ref{sec:frame}, we present our unifying framework for slow driving in non-Hermitian systems.  Applying it to models of dissipative qubits, we show in Secs. \ref{sec:NHH} and \ref{sec:lind}, that the framework consistently captures CSC under NHH, Lindblad and hybrid-Lindblad evolution.  Finally, in Sec. \ref{sec:glo}, we discuss CSC in a non-Hermitian model with no EPs.

\section{Preliminaries}
\label{sec:prelim}

\subsection{Hybrid-Lindblad evolution}

Throughout this work, we consider scenarios in which a system evolves according to the hybrid-Lindblad master equation \cite{Minganti2020} ($\hbar = 1$)
\begin{align}\label{eq:lindME}
    \dot{\underline \rho} = \mathcal L_q(t)\underline\rho = -i( H_{\text{eff}}\underline\rho- \underline\rho   H_{\text{eff}}^\dagger) + q \sum_i\gamma_i L_i\underline \rho L_i^\dagger,
\end{align}
whose solution gives the time-evolved state
\begin{align}\label{eq:Lqstate}
    \underline\rho(t) = \mathcal T e^{\int_0^t dt' \mathcal L_q(t')dt'}\rho(0).
\end{align}The underbar indicates that the state is in general not normalised; the normalised state is given by $\rho(t)=\underline \rho(t)/\tr(\underline \rho(t))$. We have defined an effective non-Hermitian Hamiltonian (NHH) $ H_{\text{eff}} \coloneqq H-\frac{i}{2}\sum_i \gamma_iL_i^\dagger L_i$, where $H$ is the system Hamiltonian and $L_i$ are the jump operators with associated rates $\gamma_i$. The first term in Eq.~\eqref{eq:lindME} induces coherent non-unitary loss, and quantum jumps in the second term represent the continuous monitoring of the system by the environment \cite{Wiseman}. The parameter $q \in [0,1]$ in the hybrid-Liouvillian $\mathcal L_q$ allows us to interpolate between full quantum and purely NHH dynamics. Specifically, $q=1$ corresponds to the standard Lindblad equation; we denote the corresponding Liouvillian superoperator by $\mathcal L\equiv \mathcal L_1$, and it is the generator of completely positive and trace-preserving evolution of the system. On the other hand, $q = 0$ corresponds to pure NHH dynamics, and represents evolution with quantum jumps completely postselected out. The case $0<q<1$ can be interpreted as the result of imperfect postselection, with $1-q$ being the degree of postselection. Importantly, the superoperator $\mathcal L_q$ is not trace-preserving for $0\leq q<1$, i.e., the resulting time-evolved state $\underline\rho(t)$ requires renormalisation. For numerical calculations with Eq. \eqref{eq:Lqstate}, we discretise the time-ordered exponential and renormalise the state at every time step. 

In particular, we will be interested in pure NHH evolution with $q=0$ and full Lindblad evolution with $q=1$. The Lindblad equation, except for certain simple cases such as spontaneous decay to a ground state or anomalous cases, e.g., the production of dark states \cite{Khandelwal2024}, leads to an unavoidable change in the purity of a quantum state. However, if the initial state is pure, NHH evolution preserves its purity \cite{Zloshchastiev2015,Khandelwal2024a}. In this case, the evolution of an unnormalised state can be expressed in the form of a non-Hermitian Schrödinger equation,
\begin{align}\label{eq:nlse}
  |\underline{ \dot\psi}(t)\rangle = -i H_{\text{eff}}| \underline \psi (t)\rangle.
  \end{align}
If the initial state is mixed, one must resort to the more general form, $\underline{\dot \rho}(t) = \mathcal L_0 \underline\rho(t)$.

\subsection{Adiabatic evolution: Hamiltonian, non-Hermitian Hamiltonian and Lindblad cases}

The standard adiabatic theorem in Hermitian quantum mechanics states that, under slow variation of time-dependent parameters, a system in an eigenstate of its Hamiltonian will remain in the same instantaneous eigenstate. This prevents any jumps or switching between Hamiltonian eigenstates under adiabatic evolution. However, this is known not to be the case for dissipative systems subject to NHH evolution. The key difference between adiabatic theorems for Hermitian and non-Hermitian Hamiltonians lies in the fact that the latter have a complex eigenvalue spectrum. 

In the non-degenerate and diagonalisable NHH case, a straightforward generalisation of the Hermitian adiabatic theorem can be written in the following way \cite{Garrison1988},
\begin{equation}\label{eq:adnhh}
|\underline \psi(t)\rangle_{\text{ad}} = \sum_i c_i(0) e^{-i\int_0^t \xi_i(t')dt'} e^{-\int_0^t \langle \phi_i|\dot{\psi}_i\rangle dt'} |\psi_i(t)\rangle,
\end{equation}where $\ket{\psi(0)}=\sum_i c_i(0)\ket{\psi_i}$. Here, $|\underline \psi(t)\rangle_{\text{ad}}$ is the unnormalised time-evolved state in the adiabatic limit. The left and right eigenstates of the NHH are defined by $\langle \phi_i|  H_{\text{eff}} = \xi_ i \langle \phi_i|$ and $ H_{\text{eff}} |\psi_i\rangle = \xi_i \ket{\psi_i}$, such that $\langle \phi_i |\psi_j \rangle = \delta_{ij}$. Due to the complex eigenvalue spectrum, the factors in the exponential terms cannot be interpreted as pure phases as in the standard adiabatic theorem. Instead, the complex part of the eigenvalues can result in an exponential growth or decay of certain eigenstates. This may lead to one eigenstate dominating the above decomposition at long times, while the others vanish. Depending on the setup and parameters, almost any initial state is then led to the most growing eigenstate, contrary to the usual adiabatic prediction. However, the adiabatic expression \eqref{eq:adnhh} ignores coupling between different eigenstates, which means that there cannot be any conversion between them. This is at odds with both theoretical prediction and experimental observation \cite{Kumar2021,Sun2023,Khandelwal2024a,Chen2022}. The simple reason for this apparent discrepancy is that no trajectory, whether numerically or experimentally, is truly adiabatic. Non-adiabatic corrections to the evolution must then couple different eigenstates and allow state conversion. This necessitates building a theory for slow non-Hermitian evolution. 

The full Lindblad equation presents significant differences. First, the almost unavoidable loss of purity under Lindblad evolution implies that there can never be perfect state conversion in this case; if the system starts in a pure state, it can essentially never evolve to another pure state. Nevertheless, state conversion can happen with a reduced fidelity \cite{Kumar2021,Khandelwal2024a}. Second, finite-dimensional Liouvillians have a steady state. This means that in the adiabatic limit, where the driving time is considerably larger than the relaxation time of the dynamics, the system reaches the steady state at every point along the trajectory. Thus, we must have the adiabatic state $\rho_{\text{ad}}(t) \simeq \rho_{\text{ss}}^t$, where $\rho_{\text{ss}}^t$ is the steady state of $\mathcal L(t)$. In such a case, we cannot expect state conversion as the steady state is independent of the orientation of the trajectory and dependent only on the zero eigenmatrix of the Liouvillian.  This implies that state conversion in the Lindblad case is necessarily a non-adiabatic effect, and an appropriate description requires non-adiabatic corrections to the adiabatic (steady-state) prediction.

\subsection{Limiting cases}\label{sec:lim}

Without resorting to the adiabatic theorem, there are two simple limiting cases of NHH dynamics \eqref{eq:nlse} that can be dealt with straightforwardly and exactly, without reference to the system in question. The first is when dissipation is uniform, i.e., when the NHH is of the form $H_{\text{eff}} = H-i\gamma\mathds 1$, where $\gamma$ is a coupling parameter. The normalised time-evolved state can be expressed as
\begin{equation}
	\begin{aligned}
\ket{\psi(t)} &=  \frac{\mathcal{T} e^{-i  \int_0^t H(t')-i\gamma(t') \mathds 1 dt'} |\psi (0)\rangle }{ \| \mathcal{T} e^{-i \int_0^t H(t')-i\gamma(t') \mathds 1 dt'} |\psi (0)\rangle\|_2 } \\
&=\mathcal{T} e^{-i \int_0^t H(t') t'} |\psi (0)\rangle
\end{aligned}
\end{equation}Here, we have used that all operators commute with the identity to extract the dissipation term outside the time-ordering, and the norm-preserving property of the Hamiltonian. Therefore, for equal dissipation rates, the normalised time-evolved state under NHH evolution is identical to the Hermitian Hamiltonian case. In accordance with the standard adiabatic theorem, if the system starts in an eigenstate of the Hamiltonian, under slow variation of parameters it will remain in the same instantaneous eigenstate. Therefore, there can be no state conversion with uniform dissipation rates. 

The second simple case when there is no coupling in the NHH, i.e, it is diagonal in the energy eigenbasis. Then it takes the simple form $H_{\text{eff}} = H - i\Gamma$, where $[H]_{jj}= \epsilon_j(t)\ketbra{j}{j}$ and $[\Gamma]_{jj}= \gamma_j(t)\ketbra{j}{j}$. Since the NHH is diagonal, it commutes at different times, $[H_{\text{eff}}(t),H_{\text{eff}}(t')]=0$. This implies that time-ordering is not necessary and can be ignored. Moreover, the computational basis states $\{\ket{j}\}_j$ are eigenstates of the NHH at all times. Let the initial state $\ket{\psi(0)} = \sum_j c_j(0)\ket{j}$ evolve under a NHH. Then at a later time $t$ we have the normalised state
\begin{equation}
\begin{aligned}
|\psi(t)\rangle = \frac{\sum_j c_j(0)e^{-i\int_0^t \epsilon_j(t')dt'}e^{-\int_0^t \gamma_j(t')dt'}\ket{j}}{\sqrt{\sum_j |c_j(0)|^2 e^{-2\int_0^t\gamma_j(t')dt'}}}
\end{aligned}
\end{equation}Therefore, up to a phase, only the least decaying state will survive. In particular, if we start in an eigenstate $\ket{j^*}$, $c_{j^*}(0)=1$ (and $c_j(0)=0$ for $j\neq j^*$), we obtain
\begin{equation}
	\ket{\psi(t)}=e^{-i\int_0^t \epsilon_{j^*}(t')dt'}\ket{j^*} \simeq \ket{j^*}.
\end{equation}Starting in an eigenstate, we remain in the same state up to an unimportant global phase factor, making state conversion impossible. 

Therefore, there are two essential requirements for state conversion: a disbalance between the dissipation rates and coupling within the system Hamiltonian. These conditions will be naturally seen in the examples considered below.

\subsection{Models}

To demonstrate our results, we will consider the following models in this work. Numerical calculations with model A will be presented in the main text, while ones for model B can be found in the appendix.

\subsubsection{Model A: Single dissipative qubit}

Model A consists of a qubit weakly coupled to a thermal reservoir. The Hamiltonian is $H^{\text{A}} =\delta \sigma_+\sigma_- + \kappa\sigma_x$, where $\sigma_\pm$ are ladder operators for the qubit, $\delta$ is the splitting between the energy levels, and $\kappa$ is a tunnel coupling. The evolution can be described with the following hybrid-Lindblad equation
\begin{align}
    \dot{\underline{\rho}} = -i\left(H^{\text{A}}_\text{eff}\underline\rho-\underline\rho H^{\text A\dagger}_{\text{eff}}\right) + q(\gamma^+\sigma_+\underline\rho\sigma_- + \gamma^- \sigma_-\underline\rho\sigma_+),
\end{align}where $\gamma^+$ and $\gamma^-$ are excitation and de-excitation rates. A second-order EP in $H^{\text{A}}_\text{eff}$ is located at $\{\delta = 0,\kappa = \pm (\gamma^- -\gamma^+) /4\}$. CSC has been shown to occur in similar models \cite{Kumar2021,Chen2022} between the states $\ket{\pm}=(\ket{0}\pm\ket{1})/\sqrt{2}$, which are eigenstates of $H^{\text A}$ for $\delta=0$.

\subsubsection{Model B: Coupled dissipative qubits}
Model B consists of two coupled qubits (denoted by $k=1,2$), each weakly interacting with a thermal reservoir with bare coupling $\gamma_k$. The Hamiltonian is $H^{\text{B}} = \sum_{k=1,2} \varepsilon_k \sigma^{\left(k\right)}_+\sigma^{\left(k\right)}_- + g\left(\sigma^{\left(1\right)}_+\sigma^{\left(2\right)}_- +\sigma^{\left(1\right)}_-\sigma^{\left(2\right)}_+\right)$, with transition energies $\varepsilon_1 = \varepsilon$ and $\varepsilon_2 = \varepsilon+\delta$. Assuming weak inter-qubit interaction such that $g\lesssim\gamma_k\ll \varepsilon_k$, the dynamics of the two qubits are governed by a local hybrid-Lindblad equation of the form \cite{Hofer2017, Khandelwal2024a}
\begin{equation}
\label{eq:lind}
\begin{aligned}
	\dot{\underline{\rho}} 
	=  &-i\left(H^{\text{B}}_\text{eff}\underline\rho-\underline\rho H^{B\dagger}_{\text{eff}}\right) \\&+ q\sum_{k=1,2} \gamma_k^+ \sigma^{(k)}_+ \underline\rho \sigma_{-}^{(k)} + \gamma_k^- \sigma^{(k)}_{-} \underline\rho \sigma_{+}^{(k)}\,,
\end{aligned}
\end{equation}
where $\sigma^{\left(k\right)}_\pm$ are the raising and lowering operators corresponding to qubit $k$. The associated excitation and de-excitation rates for qubit $k$ are denoted by $\gamma_k^+$ and $\gamma_k^-$, respectively. Of particular interest for CSC \cite{Khandelwal2024a} is the case with gain only on one qubit and loss only on the other. Therefore, we will set $\gamma_1^-=\gamma_2^+=0$ throughout.   A second-order EP in $H^{\text{B}}_\text{eff}$ lies at $\{\delta = 0,g = (\gamma_1^++\gamma_2^-)/4\}$. CSC has been shown to occur between the Bell states $\ket{\Psi^\pm}=(\ket{01}\pm\ket{10})/\sqrt{2}$ \cite{Khandelwal2024a}, which are eigenstates of $H^{\text B}$ for $\delta=0$.

\subsection{Trajectory and figure of merit for CSC}

 Throughout this work, we choose the following periodic driving of the parameters for calculations with the above models,
\begin{equation}\label{eq:traj_mod}
\delta(t) =   \delta_0 \sin(2 \pi x \frac{t}{T}), \quad 
\gamma(t) = \gamma^\prime + \gamma_0 \sin^2(\pi \frac{t}{T}).
\end{equation}
Here, $T$ is the period, $\gamma^\prime$ sets the origin and $ \delta_0$ and $ \gamma_0$ are the amplitudes for $\delta(t)$ and $\gamma(t)$, respectively. The sign $x = \pm 1$ corresponds to clockwise (CW) and counter-clockwise (CCW) trajectories, respectively. Similar trajectories have been commonly used in literature in the context of CSC \cite{Kumar2021, Sun2023,Chen2022,Khandelwal2024a,Kumar2025}. 

To quantify state conversion along the trajectory, we consider the fidelity between a suitable target state and the time-evolved state $\rho(t)$. In general, the target state is pure $\ket{\psi^{\text{targ}}}$. Then the fidelity reads 
\begin{equation}
F(\psi^{\text{targ}},\rho(t)) = \langle \psi^{\text{targ}}|\rho(t)|\psi^{\text{targ}} \rangle.
\end{equation}Let us also note that evolution with non-Hermitian operators is often noisy and exhibits fast oscillations. Therefore, in these scenarios, an alternative choice can be to work with a moving average of the fidelity in time, rather than the fidelity directly.

\section{Unified framework for slow driving}
\label{sec:frame}

We consider a quantum state $\rho(t)$ evolving under the action of some time-dependent superoperator $\mathcal{S}_t$ (where the subscript $t$ denotes time dependence) with slowly varying parameters, $\partial_t \underline\rho(t) = \mathcal{S}_t \underline\rho(t)$. For example, $\mathcal S_t$ could correspond to the Hamiltonian, Liouvillian, or hybrid-Liouvillian superoperator. Therefore, $\mathcal S_t$ may be non-Hermitian or even non-trace-preserving. Let $T$ be the total time interval during which the system evolves under the influence of the external control and let $s = t/T \in [0,1]$ be the rescaled time. The rescaled equation of motion is given by 
\begin{equation}
\partial_s \underline\rho(s) = T\mathcal{S}_s \underline\rho(s).
\end{equation} We assume that $\mathcal S_s$ is diagonalisable for all $s\in[0,1]$, i.e., the parameter trajectory does not cross any exceptional points. We further assume that the eigenspectrum remains non-degenerate throughout the evolution, which is necessary in the standard formulation of adiabatic evolution. In general, the eigenvalues of $\mathcal S_s$ are complex and there exist distinct right and left eigenmatrices $\rho_i$ and $\sigma_i$, defined by $\mathcal{S}_s \rho_i = \lambda_i \rho_i$ and $\sigma_i^\dagger \mathcal{S}_s = \lambda_i \sigma_i^\dagger$. They are bi-orthogonal with respect to the Hilbert-Schmidt inner product, and can be chosen to be normalised such that $\tr(\sigma_i^\dagger \rho_j) = \delta_{ij}$. For convenience, we have omitted time dependence from the eigenvalues and eigenvectors; it is important to note that it is implicitly there.

We assume that the time scale of the externally controlled parameters is much longer than the characteristic relaxation time of the system. In App. \ref{App:HH}, we derive the first non-adiabatic correction for arbitrary $\mathcal S_s$ and show that the unnormalised time-evolved state in vectorised notation ($\rho\to\kett{\rho}$) can be expressed as 
\begin{align}
    \kett{\underline \rho(s)} \approx \kett{\underline \rho_{\text{sd}}(s)} = \mathds{U}(s,0)\kett{\rho(0)},
\end{align}where the subscript ``sd" stands for slow-driving. The normalised state is given by $\kett{\rho(s)} = \kett{\underline\rho(s)}/\brakett{\mathds 1}{\underline\rho(s)}$, where the denominator is the trace of $\underline\rho(s)$ in vectorised notation, with $\kett{\mathds{1}}$ the vectorised identity matrix. The slow-driving evolution operator $\mathds{U}(s,0)$ induces slow evolution of the system, including the first non-adiabatic correction. Below, we write the forms of $\mathds{U}(s,0)$ for the cases relevant for this work. More details can be found in App. \ref{App:HH}. 

Let us note that although we have assumed trajectories not passing through EPs, it is possible to relax this assumption and derive general results \cite{Sarandy2005}. The key difference will be the non-diagonalisability of the involved superoperators, which will require the construction of a Jordan form and generalised eigenvectors. However, recent work has numerically shown no significant difference in CSC with such trajectories in either NHH or Lindblad cases \cite{Khandelwal2024a}. 

Lastly, it is possible to derive perturbative corrections to arbitrary orders \cite{Rigolin2008,Cavina2017,Kumar2025}; we discuss these in detail in App. \ref{App:HH}. In our examples, we find that the lowest order is sufficient to capture the essence of CSC.

\subsection{Hybrid-Lindblad evolution}
\label{sec:hybsd}
In the case of hybrid-Lindblad evolution ($0\leq q\leq1$), $\mathcal S_s = \mathcal L_q(s)$. The evolution operator takes the form 
\begin{equation}\label{eq:NHHU}
\begin{aligned}
&\mathds{U}_q(s,0) = \sum_k e^{T \int_0^s \ ds' \Delta^q_{kk} (s')} \ketbraa{\rho^q_k(s)}{\sigma^q_k(0)}\\
&+ \frac{1}{T}\sum_{m \neq k} e^{T \int_0^s \ ds' \Delta^q_{kk}(s') } \frac{\brakett{\sigma^q_m(s)}{\dot{\rho}^q_k(s)}}{\lambda^q_m(s)-\lambda^q_k(s)} \ketbraa{\rho^q_m(s)}{\sigma^q_k(0)}\\
&- \frac{1}{T}\sum_{m \neq k} e^{T \int_0^s \ ds' \Delta^q_{kk}(s') } \frac{\brakett{\sigma^q_k(0)}{\dot{\rho}^q_m(0)}}{\lambda^q_k(0)-\lambda^q_m(0)}\ketbraa{\rho^q_k(s)}{\sigma^q_m(0)}.
\end{aligned}
\end{equation}
The exponents $\Delta^q_{kk}$ are given by 
\begin{equation}\label{eq:decay_raye}
\Delta^q_{kk} = \lambda^q_k - \frac{1}{T} \brakett{\sigma^q_k}{\dot{\rho}^q_k}.
\end{equation}
We refer to $ \tilde \Delta_k (s) \equiv \int_0^s ds' \, \Delta^q_{kk}(s')$ as a growth parameter. In general, the evolution operator $\mathds{U}_q(s,0)$ evolves any initial state $\kett{\rho(0)}$ towards the instantaneous eigenstate $\kett{\rho^q_{k}}$ of the superoperator $\mathcal L_q$ with the largest growth parameter. 

Let us note some connections to past works. First, if there is no dissipation (Hamiltonian evolution), this procedure is a generalisation of the Hermitian adiabatic perturbation theory of Ref. \cite{Rigolin2008} for mixed states. For non-zero dissipation and in the case $q=0$, we obtain a NHH version of the same. This case has been considered in Ref. \cite{Kumar2025} for pure states.

\subsection{Lindblad evolution}

We consider the Lindblad ($q=1$) case separately as it leads to significant simplification and gives elegant analytical results. In this case, $\mathcal{S}_s = \mathcal{L}(s)$. Finite dimensional Liouvillians have at least one steady state \cite{Rivas2012}. We assume that at every point along the parameter trajectory, the Liouvillian has a unique steady state and adopt the convention $\lambda_0 = 0$ so that $\dket{\rho_0} = \dket{\rho_{\text{ss}}}$ and $\dbra{\sigma_0} = \dbra{\openone}$ \cite{Minganti2018}. Note that if the Liouvillian has a unique steady state, the exponent $e^{ T \int_0^1 ds' \Delta(s')} \simeq \dketbra{0}{0}$ for large $T$. The slow-driving evolution operator then takes the form
\begin{equation}\label{eq:LU}
\begin{aligned}
&\mathds{U}(s,0) \\&= \left(\kett{\rho_{\text{ss}}(s)} + \frac{1}{T}\sum_{m \neq 0} \frac{\dbraket{\sigma_m(s)}{\dot{\rho}_{\text{ss}}(s)}}{\lambda_m(s)} \kett{\rho_m(s)}\right)\braa{\openone}
\end{aligned}
\end{equation}
The zeroth-order (adiabatic) prediction is the instantaneous steady state of $\mathcal L(s)$, while the correction depends only on the properties of $\mathcal L(s)$ and of the parameter trajectory. Moreover, unlike the hybrid-Lindblad case, the slow-driving evolution operator is independent of the initial state. Eq. \eqref{eq:LU} is equivalent to the slow-driving perturbation theory for the Lindblad equation developed in Ref. \cite{Cavina2017}.

\subsubsection{Connection to Floquet-Lindblad theory}
\label{sec:flo}

Since our examples will involve periodic driving, it is interesting to apply Floquet theory, the standard approach for periodically-driven systems. Although Floquet theory has mostly been used for Hamiltonian evolution in closed quantum systems, recent work has revealed that Floquet solutions can also be constructed for open quantum systems under Lindblad evolution \cite{Scopa2018,Sato2025}, in spite of some ambiguities regarding the existence of a Floquet Liouvillian \cite{Schnell2020}. Moreover, high-frequency solutions to such cases have been given \cite{Mizuta2021,Schnell2021} with methods inspired from Hamiltonian Floquet theory. However, here we are concerned with the low-frequency or slow-driving case, which should lead us to chiral state conversion under the right conditions.

We have determined the slow-driving evolution operator for Lindblad evolution $\mathds U(t,0) $ in Eq. \eqref{eq:LU}. Therefore, the slow-driving one-period or Floquet evolution operator is
\begin{align}\label{eq:floev}
     \mathds U_F \equiv \mathds U(T,0) = \kett{\rho_{\text{sd}}(T)}\braa{\mathds 1}.
\end{align}Acting this operator onto any initial state $\kett{\rho(0)}$, we obtain the slow-driving prediction $\kett{\rho_{\text{sd}}(T)}$. We therefore have that the final state $\kett{\rho_{\text{sd}}(T)}$ is precisely the 1-eigenvector of $\mathds U_F$ and corresponds to the Floquet steady state. Naturally, for any $t=nT+\tau$ ($n\in \mathds N$), we have $\mathds U(t,0)=\mathds U(\tau,0)\mathds U_F$. 

Having the Floquet evolution operator at hand, we can comment on the existence of a Floquet Liouvillian $L_F$, defined through $\mathds U_F = e^{TL_F}$. In finite dimensions and for finite $t$, the evolution operator $e^{Lt}$ corresponding to any generator $L$ is an invertible map. However, the Floquet evolution operator \eqref{eq:floev} has a unique non-zero eigenvalue (equal to 1) and is clearly non-invertible. This means that there cannot exist a Floquet Liouvillian corresponding to $\mathds U_F$, for finite driving period $T$.

Lastly, we note that $\mathds U_F$ corresponds to a completely positive and trace-preserving reset quantum channel $\mathcal U_F$  with $\rho_{\text{sd}}(T)$ as the target of the reset. That is, for any given state $\rho(0)$, we have $\,\mathcal U_F(\rho(0)) = \rho_{\text{sd}}(T)$. If the spectral decomposition of the target takes the form $\rho_{\text{sd}}(T)=\sum_i{p_i}\ketbra{i}{i}$, then the reset Lindblad generator can be written as $ \mathcal L_{\text{res}}(\rho) = \sum_{jk}\gamma_{jk}(L_{jk}\rho L^\dagger_{jk} - \{L^\dagger_{jk}L_{jk},\rho\}/2)$, where $L_{jk}=\ketbra{k}{j}$ and the rates satisfy $\gamma_{jk}/\gamma_{kj}=p_k/p_j$. It can be checked that the steady state of $\mathcal L_{\text{res}}$ is precisely $\rho_{\text{sd}}(T)$. However, this construction by definition corresponds to the Floquet Liouvillian only in the trivial limit $T\to\infty$, in which the system ends up in the instantaneous steady state at every point along the parameter trajectory.

\section{Chiral state conversion under hybrid-Lindblad dynamics}
\label{sec:NHH}

\subsection{NHH evolution}

\label{sec:sdNHH}

We give brief solutions here; details can be found in App. \ref{App:CSC_NHH}.

\subsubsection{Model A}

\begin{figure*}%[ht!]
	\centering
	\includegraphics[width=0.85\textwidth]{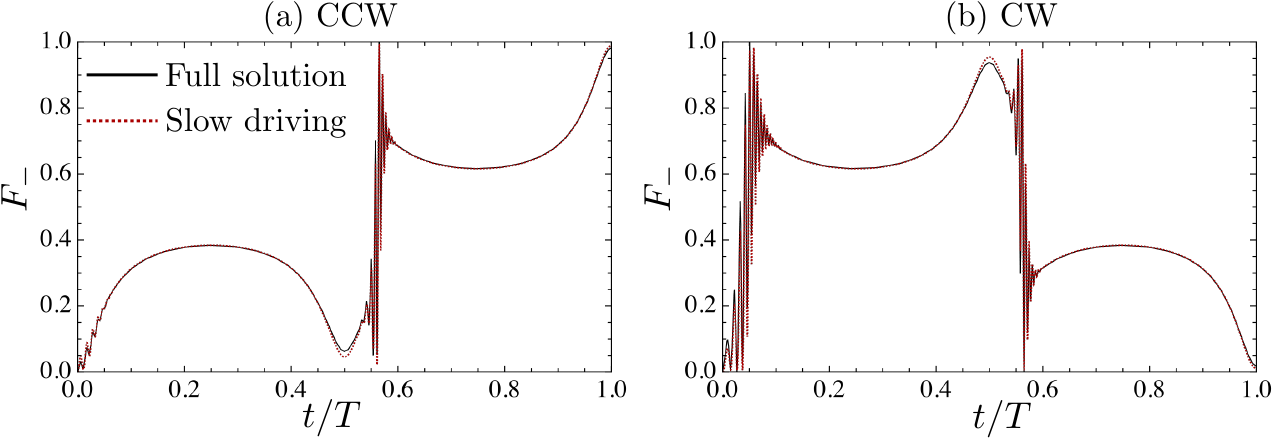}
	\caption{Conversion fidelity $F_-$ as a function of rescaled time $s= t/T$ for (a) CCW and (b) CW trajectories with the initial state $\rho(0)=\ketbra{+}{+}$, with $\gamma^+=0$ and $\gamma^-=\gamma(t)$.  The slow-driving prediction corresponds to a two-step application of the slow-driving operator with $s^*=1/2$. Parameters: $\delta_0=1$, $\gamma'=0.1\delta_0$, $\gamma_0=0.1\delta_0$, $\kappa=0.12\delta_0$ and $\gamma' T =200 $.
    }\label{Fig_1qub_Heff}
\end{figure*}
We start by considering model A, with the energy splitting $\delta(s)$ and coupling rate $\gamma(s)$ given by Eq.~\eqref{eq:traj_mod}. For generality, we consider an arbitrary mixed initial state $\rho(0)$; the analysis holds for pure initial states as well. The instantaneous eigenvalues of the NHH take the form,
\begin{align}
   \xi_{\pm}(s) = \frac{1}{4}\left(2\delta-i\Gamma^+\pm \eta \right),
\end{align}where $\Gamma^{\pm} = \gamma^-\pm \gamma^+$ and $\eta = \sqrt{16\kappa^2+(2\delta-i\Gamma^-)^2}$. The corresponding unnormalised eigenstates are
\begin{align}
\ket{\underline{\psi}{}_{\pm}} = %\frac{1}{\sqrt{1+\frac{(\gamma-2i\lambda_\pm)(\gamma+2i\lambda^*_\pm)}{4\kappa^2}}}
\ket{0}+\frac{2\xi_\pm+i\gamma}{2\kappa}\ket{1} 
\end{align}We have omitted time dependence from parameters for notational convenience. As $H_{\text{eff}}^A$ is symmetric, the left eigenvectors are simply $\ket{\underline{\phi}{}_i}=\ket{\underline\psi{}_i}^*$. The normalised eigenstates are given by $\ket{\psi_\pm} = \ket{\underline{\psi}{}_{\pm}}/\sqrt{\braket{\underline{\phi}{}_{\pm}|\underline{\psi}{}_{\pm}}}$. Note that in the special case $\gamma' = 0$, the eigenvectors are $\{\ket{+}, \ket{-}\}$ at the initial and final points of the trajectory. The eigenvalues of $\mathcal L^A_0$, $\{\lambda^0_{lm}\}_{lm}$, and eigenvectors, $\{\rho^0_{lm}\}_{lm}$ ($l,m\in\{+,-\}$), are related to those of $H_{\text{eff}}^{\text{A}}$ through the equations $\lambda^0_{lm} = -i(\xi_m -\xi_l^*)$ and $\rho^0_{lm} = \ketbra{\psi_l}{\psi_m}$ \cite{Minganti2019}. Importantly, only two of the eigenvectors, $\rho^0_\pp$ and $\rho^0_\mm$ are valid quantum states under normalisation; these are the key states for CSC. 

Within slow driving, the evolution is described by the operator $\mathbb U^{\text{A}}_0$, given in Eq. \eqref{eq:NHHU}. Our framework predicts that the vectorised initial state $\kett{\rho(0)}$ will evolve towards the instantaneous right eigenstate $\kett{\rho^0_\pp}$ or $\kett{\rho^0_\mm}$ which has a larger corresponding growth parameter $\tilde \Delta^0_{\text{\scriptsize{$\pm\pm$}}}(s) = \int_0^s \, ds' \lambda^0_{\text{\scriptsize{$\pm\pm$}}}(s') - \frac{1}{T}\brakett{\sigma^0_{\text{\scriptsize{$\pm\pm$}}}(s)}{\dot \rho^0_{\text{\scriptsize{$\pm\pm$}}}(s)}$. For large $T$, approaching adiabaticity, the contribution in the growth parameter from the second term is negligible and 
\begin{align}\label{eq:dec1}
    \tilde \Delta^0_{\text{\scriptsize{$\pm\pm$}}}(s) \approx \int_0^s \, ds' \lambda^0_{\text{\scriptsize{$\pm\pm$}}}(s').
\end{align}

We note three key features of the growth parameters; see App. \ref{App:CSC_NHH1} for details. First, normalisation of the state at the end of the trajectory removes any contribution from the $\Gamma^+$ terms in the eigenvalues. Therefore, these terms can be safely neglected. Second, for uniform dissipation rates ($\gamma^+=\gamma^-$), it can be shown that the growth parameters are equal $\tilde \Delta^0_{\pp}(s)=\tilde \Delta^0_{\mm}(s)$. This, in turn implies that there can be no state conversion, as expected.  Third, and most importantly, the real part of the eigenvalues is odd-symmetric around $s=1/2$. At the final point of trajectory, this leads to the time integral over the real part vanishing, $\Re( \tilde \Delta^0_{\text{\scriptsize{$\pm\pm$}}}(T)) =0$. Hence, Eq.~\eqref{eq:adnhh} fails to predict state conversion for the trajectory \eqref{eq:traj_mod}. This is in contradiction with numerical \cite{Kumar2021}  and experimental \cite{Chen2022} observation. The failure of slow dynamics to capture state conversion in this model has been noted in Ref. \cite{Kumar2025}. The solution therein is to introduce fast noise that drives state conversion in the system. However, we approach the solution from a phenomenological perspective, which leads to quantitatively correct predictions and does not require any additions to the slow-driving framework.

First, we note that due to the form of the trajectory, both growth parameters $\tilde \Delta^0_{\pm\pm}$ lead to growth and decay for half the trajectory. The consequence is that over one full trajectory, the most growing (or least decaying) state at the end is the same one as the beginning. As a result, applying the slow-driving approach over the whole trajectory directly cannot lead to state conversion. Second, the full solution consists of a time-ordered exponential that is solved by discretising the evolution and renormalising the state at every time step. This ``resetting" of the state is not captured by a direct application of the slow driving operator over the entire trajectory, followed by renormalisation. Taking into account these points, the natural solution to this problem lies in splitting the application of the slow-driving operator into two time intervals. The evolution of the system from $s=0$ to some $s\leq s^*$ is governed by $\kett{\underline\rho(s)} = \mathbb U^{\text{A}}(s,0)\kett{\rho(0)}$. At $s=s^*$, the system is renormalised and the evolution beyond this point is governed by the operator $\mathbb U^{\text{A}}_0(s,s^*)$,
\begin{align}
\kett{\underline\rho(s)} = \mathbb U^{\text{A}}_0(s,s^*) \frac{\kett{\underline\rho(s=s^*)}}{\brakett{\mathds 1}{\underline\rho(s=s^*)}},
\end{align}where $\kett{\underline\rho(s=s^*)}=\mathbb U^{\text{A}}_0(s^*,0)\kett{\rho(0)}$. The state $\kett{\underline\rho(s)}$ is again renormalised with its trace $\brakett{\mathds 1}{\underline\rho(s)}$. In this two-step slow-driving procedure, the growth parameter that determines the dominant state is no longer \eqref{eq:dec1}, but is rather 
\begin{align}
     \tilde \Delta^0_{\text{\scriptsize{$\pm\pm$}}}(s) \approx \int_{s^*}^s \, ds' \lambda^0_{\text{\scriptsize{$\pm\pm$}}}(s').
\end{align}Then, it can be shown that for a CW (CCW) trajectory, the larger growth parameter at the end of the trajectory is $\tilde \Delta^0_\pp(T)$ $(\tilde \Delta^0_\mm(T))$, evaluated with the above equation. Hence, any initial state $\rho(0)$ will evolve towards the associated eigenstate.

In Fig. \ref{Fig_1qub_Heff} we plot the conversion fidelity $F_-(s) = \bra{-}\rho(s)\ket{-}$ as a function of the rescaled time $s=t/T$, with the initial state $\rho(0)= \ketbra{+}{+}$. As described above, we have used a two-step application of the slow-driving evolution operator, taking $s^*=1/2$. The slow-driving prediction (red) captures the full dynamics (black) governed by time-ordered evolution operator for essentially all times. Moreover, we even find that the noisy regions obtained with the two approaches show a considerable overlap. We further discuss the role of the choice of $s^*$ in App. \ref{App:CSC_NHH1}.

Let us briefly note the role of the EP; further details can be found in App. \ref{App:CSC_NHH1}. As mentioned previously, the EP is located at $\{\delta=0,\gamma=4\kappa\}$. When $\kappa<\gamma'/4$, the trajectory does not encircle the EP. In this case, the system is in an underdamped regime towards the end of the trajectory, where the growth parameters become imaginary. Therefore, the slow-driving framework predicts no state conversion in this scenario. However, whenever $\kappa>\gamma'/4$, the growth parameters are real and lead to state conversion. Importantly, when $\kappa\gg\gamma'/4$, the trajectory is again non-encircling and near-perfect conversion is predicted under slow driving. Therefore, while the EP determines where the growth parameters are real (leading to state conversion), encircling it is not essential for CSC.

\begin{figure*}
    \centering
    \includegraphics[width=0.85\textwidth]{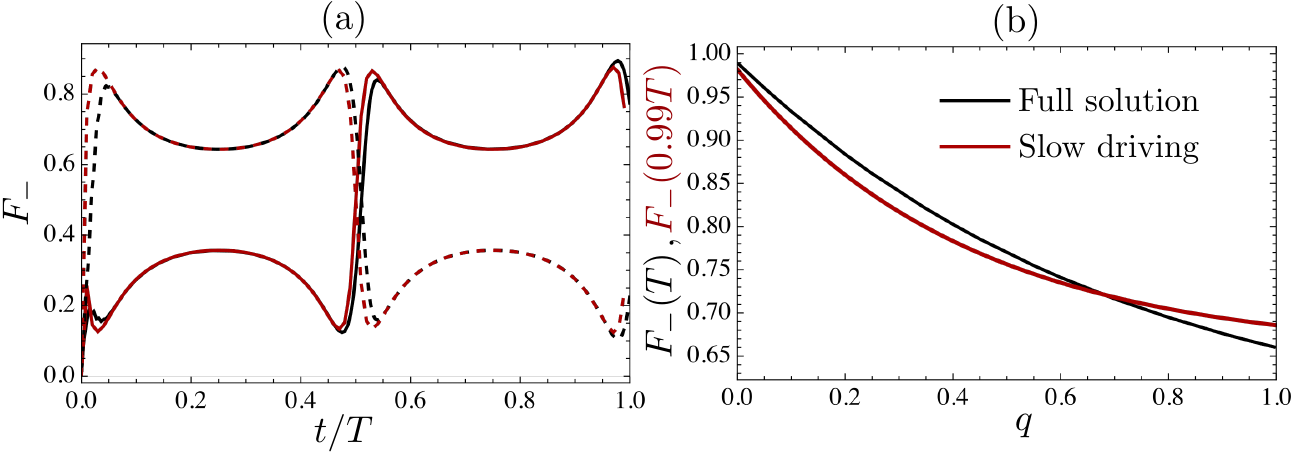}
    \caption{(a) $F_-$ as a function of $s=t/T$ for $q=0.5$ with $\rho(0)=\ketbra{+}{+}$ for CCW (solid) and CW (dashed) trajectories. The slow-driving prediction is shown up to $s=0.99$. (b) $F_-(T)$ (with the full solution) and $F_-(0.99T)$ (with the slow-driving approximation) as a function of $q$ with the same initial state and CCW trajectory. Parameters: $\gamma^+=0$, $\gamma^-=\gamma(t)$, $\delta_0=1$, $\gamma'=0.1\delta_0$, $\gamma_0=0.05\delta_0$, $\kappa=0.15\delta_0$ and $\gamma' T =200 $.}
    \label{fig:q}
\end{figure*}

\subsubsection{Model B}

We now consider model B, with the detuning $\delta(s)$ and coupling $\gamma(s)$ given by Eq.~\eqref{eq:traj_mod}. The eigenvalues of $H_{\text{eff}}^{\text{B}}$ are given by $\xi_0 = -i\gamma_1^+/2$, $ \xi_\pm = \frac{1}{4}(2\delta+4\varepsilon -i\Gamma^+\pm  \eta')$ and $ \xi_1 =-i \gamma_2^-/2+ \delta + 2\varepsilon$, where $\Gamma^{\pm} =  \gamma^+_1\pm \gamma_2^-$ and $ \eta' = \sqrt{16g^2+(2\delta-i\Gamma^+)}$. We denote the associated eigenstates by $\ket{00}, \ket{\psi_\pm}, \ket{11}$, where 
\begin{equation}
\ket{\underline{\psi} {}_\pm} = \ket{01} + \frac{2 \xi_\pm-2(\delta+ \varepsilon)+ i\Gamma^+}{2g} \ket{10},
\end{equation}
Compared to model A, the role of the coupling $\kappa$ is now played by the inter-qubit coupling $g$ and the role of the energy splitting by the energy detuning. We show in App. \ref{app:B} that only the single-excitation subspace contributes to the relevant dynamics in this model, and that the mechanism underlying CSC between the Bell states $\ket{\Psi^\pm}$ is similar to model A. However, due to the same reasons described above for model A, a one-step application of the slow-driving operator fails to capture CSC. The correct dynamics is captured by $\kett{\underline{\rho}(s)} =  \mathbb U_0^{\text{B}}(s,0)\kett{\rho(0)}$ for $s \leq s^*$, and 
\begin{align}
    \kett{\underline{\rho}(s)} =  \mathbb U_0^{\text{B}}(s,s^*)\frac{\kett{\underline{\rho}(s=s^*)}}{\brakett{\mathds 1}{\underline{\rho}(s=s^*)}},\,\,\text{for }s>s^*.
\end{align}The corresponding plot of the fidelity as a function of time can be found in App. \ref{app:B}, where we find that the slow-driving prediction matches the full solution and correctly predicts CSC.

\subsection{Hybrid-Lindblad evolution}

We now discuss CSC under hybrid-Lindblad dynamics with $q\in(0,1)$. Such a situation represents partial removal of quantum jumps from the dynamics. Hybrid evolution cannot lead to perfect state conversion as (i) the eigenstates of the hybrid-Liouvillian $\mathcal L_q$ are in general different from that of $\mathcal L_0$, and (ii) imperfect postselection leads to the loss of purity of an initially pure state. Therefore, a decrease in the conversion fidelity is expected with increasing $q$, and has been numerically seen in Refs. \cite{Kumar2021,Khandelwal2024a}. 

Let us also note a key difference between slow driving in NHH and hybrid cases. When $q>0$, we find that a one-step application of the slow-driving evolution operator manages to produce the correct dynamics. This is due to there being a non-zero real part of the relevant decay rates ($\tilde \Delta^q_{\pm\pm}$, integrated over the full trajectory) in this case, unlike the NHH case. The calculations below have therefore been performed with a one-step procedure. A two-step procedure gives similar predictions.

In Fig. \ref{fig:q} (a) for model A, we show $F_-$ as a function of $s=t/T$ for $q=0.5$. We find that the slow-driving prediction matches well with the full solution for almost the entire trajectory. However, at the end of the trajectory (i.e., very close to $s=1$), we have a large mismatch; we therefore show the curve only up to $s=0.99$. As $s\to1$, the first-order corrections to adiabatic evolution become large due to a rapid change of the eigenstates, breaking the assumption of our slow-driving framework; see App. \ref{app:hyb} for details. This means that by the end of the trajectory, non-perturbative effects completely dominate and the dynamics are no longer ``slow" by any measure. This anomaly is however highly dependent on the trajectory we have chosen and may not be present for another choice. Nevertheless, the final fidelity is captured reasonably well by the slow-driving curve. Next, in (b), we plot the final fidelity as a function of $q\in(0,1)$. For the slow-driving curve, we have calculated the fidelity at $s=0.99$. The slow-driving curve captures the monotonic decline in the fidelity and follows the full solutions closely. For these calculations, we have taken a value of $T$ compatible with slow dynamics but not approaching adiabaticity. This is to ensure that we do not trivialise state conversion as $q\to1$. However, this means that for $q>0$, higher-order corrections can play an important role. These higher-order corrections are discussed in detail in App. \ref{App:HH}. In the next section, we present the role of these corrections specifically for the $q=1$ case, where the mismatch is large but where the dynamics remain perturbative at $s=1$.

\section{Chiral state conversion under Lindblad dynamics}
\label{sec:lind}

 We again take the trajectory \eqref{eq:traj_mod} and consider the evolution of the system under full Lindblad dynamics. If the system always remains coupled to the reservoir(s), in the models that we consider, this leads to the existence of a unique steady state at every point in time. We can then directly utilise Eq.~\eqref{eq:LU} to obtain the slow-driving approximation. 

Below, we give brief analytical results for both models and numerical results for model A. More details can be found in App. \ref{app:slowdriv}.

\begin{figure*}
    \centering
    \includegraphics[width=1\textwidth]{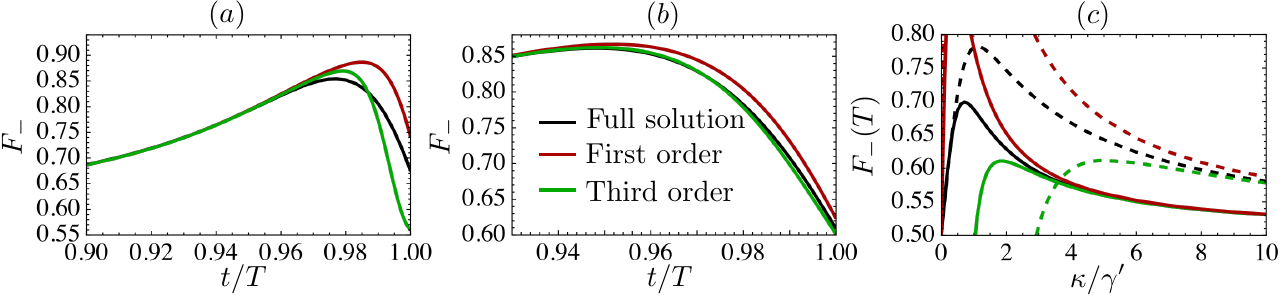}
    \caption{CSC under Lindblad evolution with $\rho(0)=\ketbra{+}{+}$ with a CCW trajectory, calculated with the full solution (black), first-order approximation (red) and second-order approximation (green). We have taken $\gamma^+=0$ and $\gamma^-=\gamma(t)$. (a) $F_-$ as a function of $s=t/T$ with parameters taken from Fig. \ref{Fig_1qub_Heff}. (b) The same except with an increased $\kappa=0.25\delta_0$. (c) $F_-(T)$ as a function of $\kappa/\gamma'$ with fixed $\gamma'=0.1\delta_0$ (solid) and $\gamma'=0.06\delta_0$ (dashed), with $\delta_0=1$ and $T=2000$.
    }
    \label{fig:slowdrivA}
\end{figure*}
\subsubsection{Model A}

To simplify analytical expressions, we assume $\gamma^+=0$ and $\gamma^-=\gamma(t)$. As before, we choose $\ket{\pm}$ as the target states for CSC. The instantaneous steady state at the end of the trajectory, $t=T$ is given by
\begin{align}
   \rho_{\text{ss}}(T)= \frac{1}{8\kappa^2+\gamma'}\begin{pmatrix}
       4\kappa^2   &-2i\kappa\gamma^\prime\\
       2i\kappa\gamma^\prime                      &4\kappa^2+\gamma'
    \end{pmatrix}. 
\end{align} Independently of the parameters of the system, the fidelity of this steady state with the target state is $F^{\text{ss}}_\pm(T)=\bra{\pm}\rho_{\text{ss}}(T)\ket{\pm}=1/2$. Therefore, the steady state has equal overlap with the targets. This reaffirms our claim - at the level of the steady state, there can be no state conversion.  Within the slow-driving approximation, the conversion fidelity at the end of the trajectory takes the form
\begin{align}\label{eq:modAF}
    F_\pm(T) = \frac{1}{2} \pm x\frac{16 \pi  \kappa\delta _0  }{\gamma'T(8  \kappa ^2 +\gamma'{}^2)}.
\end{align}The first correction is $\mathcal O(1/T)$ and is essential to state conversion. Importantly, the orientation ($x=\pm1$) of the trajectory appears in the expression for the fidelity; depending on the orientation, the fidelity can either increase or decrease. For example, the fidelity $F_+$ increases for a CW trajectory and decreases for a CCW one, while the opposite is true for $F_-$. This is the origin of the chirality of state conversion. The formula is invalid for $\gamma^\prime=0$, i.e., if the bath is decoupled from the system at the end; as this violates the assumption of our slow-driving prediction \eqref{eq:LU}. This singularity can be understood physically. The coupling determines the time scale of relaxation; for small coupling rates, the time of driving needs to be proportionally increased to counteract the increased relaxation time, maintaining $T\gamma^\prime\gg1$.

Can a better approximation to $F_\pm(T)$ be obtained at the higher orders? Interestingly, this is not always the case, as we show below. Note that for model A, the second-order correction in the expansion of the state makes no contribution to the fidelity at the end point. On the other hand, the fidelity has a $\mathcal O(1/T^3)$ correction, and its expression can be found in App. \ref{app:slowdriv}. In Fig. \ref{fig:slowdrivA} (a), we show the fidelity as a function of the rescaled time $t/T$ and a CCW trajectory. We have used the same parameters as Fig. \ref{Fig_1qub_Heff}, where we found that the slow-driving approximation perfectly captures the fidelity under NHH evolution. We obtain a nearly perfect match for intermediate times, so the plot is shown only near the end of the trajectory, where there is a mismatch.  As $s\to1$, non-adiabatic corrections evidently become large and thus the first-order approximation does not predict the fidelity perfectly, although it captures the general qualitative behaviour. Importantly, we find that the second-order approximation is worse for much of this region, clearly giving a more inaccurate prediction at $s=1$. This amplification of error suggests that the dynamics in this region are no longer slow and rather require a non-perturbative treatment. In (b), we again show the fidelity as a function of $s$, for the same parameters except $\kappa=0.25\delta_0$. Here, we find that the second-order approximation does indeed give a more accurate prediction at $s=1$; we are evidently in a perturbative regime. In Fig. \ref{fig:slowdrivA} (c), we further explore the role of the couplings in determining the perturbative regime and plot the final fidelity $F_{-}(T)$ as a function of $\kappa/\gamma'$. The plot shows that for low $\kappa/\gamma'$, we are clearly in a non-perturbative regime where the first-order approximation does not capture the fidelity accurately, and the second-order approximation amplifies the error. For larger $\kappa/\gamma'$, the second-order approximation provides a more accurate picture, indicating that the dynamics can be consistently treated perturbatively. This is rooted in the fact that for large $\kappa/\gamma'$, the first-order correction is $\mathcal O(\gamma'/\kappa)$ and therefore tends to become small. Naturally, for very large $\kappa/\gamma'$, the perturbative corrections vanish, and we approach the trivial limit $F_-(T)=1/2$.

Our numerical calculations further reveal that higher conversion fidelities can be obtained in regimes where the dynamics are non-perturbative at the end of the trajectory (as seen in Fig. \ref{fig:slowdrivA} (c)). However, in such situations, the slow-driving formula \eqref{eq:modAF} is no longer valid. Therefore, the fidelity for the two orientations is not symmetric and the conversion is not truly \textit{chiral}.

\subsubsection{Model B}
Here, to simplify analytical expressions, we assume $\gamma_1^+ = \gamma_1^-=\gamma(t)$. In analogy with model A, we consider the target states $\ket{\Psi^\pm}$. The instantaneous steady state at the end of the trajectory is given by
\begin{align}
  \rho_{\text{ss}}(T)=  \frac{ {\gamma'}^{2}}{4  {\gamma'}^{2} g^2+  {\gamma'}^{4}}\left(
\begin{array}{cccc}
 g^2 & 0 & 0 & 0 \\
 0 & \gamma'^2+g^2 & i \gamma'  g & 0 \\
 0 & -i \gamma'  g & g^2 & 0 \\
 0 & 0 & 0 & g^2 \\
\end{array}
\right),
\end{align} and the fidelity, $F_{\pm}(T)=\bra{\Psi^{\pm}}\rho_{\text{sd}}(T)\ket{\Psi^{\pm}}$, up to the first order in perturbation theory takes the form
\begin{align}\label{eq:fidtwo}
    F_{\pm}(T) = \Big(\frac{1}{2}-\frac{g^2}{4 g^2+\gamma'^2}\Big) \pm x \frac{2 \pi  g\delta_0 }{T\gamma'(4g^2+\gamma'^2)}.
\end{align}Here, at the adiabatic (zeroth-order) level, unlike model A, the fidelity is always smaller than 1/2. This is due to occupation outside the single-occupation subspace, which is always present in the steady state. The right $\mathcal O(1/T)$ term is the first correction, which leads to state conversion. It bears a clear resemblance to that of model A. The orientation of the trajectory naturally appears in the first-order correction, with target states $\ket{\Psi^\pm}$ having opposite chirality. The coupling again is key for state conversion, i.e., no state conversion for $g=0$. In the single qubit case, we found the presence of chirality also relies on detuning between the levels. Analogously, here, it relies on detuning between the qubits; there is no chirality for $\delta_0=0$. A key difference between the corrections in the two models is that model B shows a non-zero second-order fidelity correction, while model A does not.

Let us also note that model B is particularly interesting as the target states are entangled. A sufficient criterion for $\rho(T)$ to be entangled is that either $F_+>1/2$ or $F_->1/2$. As stated above, at the adiabatic level, $F_{\pm}<1/2$, which implies that the state is not necessarily entangled. The entanglement in this steady state has been characterised \cite{Khandelwal2020}. The non-adiabatic correction drives the system towards either $\ket{\Psi^+}$ or $\ket{\Psi^-}$, irrespective of the parameters. Therefore, it always tends to drive the system towards an entangled state. However, whether the state actually becomes entangled in the end depends on the involved parameters.  Moreover, $F_{+}(T,x=-1)=F_-(T,x=+1)$, which means that the amount of entanglement created, if any, is independent of the orientation of the trajectory.

We provide numerical calculations for model B in App. \ref{app:slowdriv}. When the slow-driving approximation is reasonably valid up to the first order, we find rather low conversion fidelities. This is expected due to the form of the instantaneous steady state and the occupation outside the single-occupation subspace, as explained above. However, as in model A, our analysis shows that non-perturbative behaviour at the end of the trajectory can be exploited to substantially increase the fidelity.

\section{CSC without exceptional points}
\label{sec:glo}
As we mentioned previously, encircling the EP is not essential for CSC, although its location determines whether slow dynamics can lead to state conversion. This holds for both models that we have considered. In this section, we aim to comment on whether the presence of an EP is necessary for CSC at all. To this end, we now consider model B under a global master equation approach, which corresponds to $g\gg \gamma$. In this case, the dissipators do not act locally on qubits, but rather on the joint two-qubit system, inducing transitions between eigenstates. The eigenenergies of the system Hamiltonian $H^{\text{B}}$ are $\varepsilon_0=0$, $\varepsilon_\pm= \big(\delta+2\varepsilon \pm \sqrt{4g^2+\delta^2}\big)/2$, and $\varepsilon_1 = 2\varepsilon+\delta $. The non-zero Lindblad operators are the ones corresponding to transitions of energies $\varepsilon_-$ and $\varepsilon_+$ \cite{Hofer2017}.  We denote the Lindblad jump operator of qubit $k = 1,2$ associated with transition energy $\varepsilon_\alpha$ $(\alpha = \pm)$ by $L_k(\varepsilon_\alpha)$; see App. \ref{app:globalME} for their expressions. As in model B, we assume only gain on qubit 1 and only loss on qubit 2. The global master equation then takes the form
\begin{equation}
\begin{aligned}
	\dot\rho &= \mathcal L^{\text{G}}\rho =-i\left(H^{\text{G}}_\text{eff}\rho-\rho H^{\text{G} \ \dagger}_{\text{eff}}\right)\\&+\sum_{\substack{\\\alpha\in\{-,+\}}} \gamma_1^{+}\mathcal{J}[L_1\left(\varepsilon_\alpha\right)]\rho + \gamma_2^{-}\mathcal{J}[L^\dagger_2\left(\varepsilon_\alpha\right)] \rho,
\end{aligned}
\end{equation}where the jump superoperators are defined as $\mathcal J\left[A\right]\rho\coloneqq A\rho A^\dagger$, and we have assumed energy-independent rates. Such a situation can be modeled by taking into account the exact distributions of the reservoirs \cite{Khandelwal2020}. It is known that the Liouvillian in this case does not exhibit any EPs in the dynamically relevant subspace \cite{Khandelwal2021}. We show in App. \ref{app:globalME}, that the same holds for the corresponding global NHH.

\begin{figure}
    \centering
    \includegraphics[width=0.85\columnwidth]{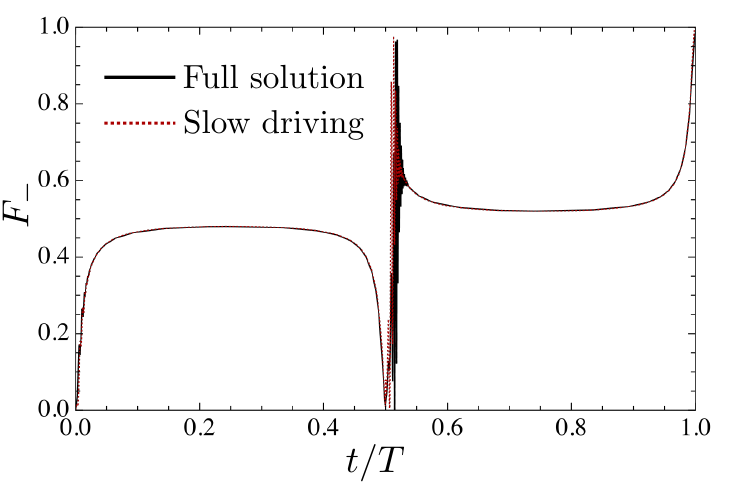}
    \caption{$F_-$ as a function of $s=t/T$ with the global master equation for $q=0$ and CCW trajectory. Parameters: $\delta_0 =10 $, $\gamma'=0.01\delta_0$, $\gamma_0=0.005\delta_0$, $g=0.02\delta_0$, $\gamma'T=200$, $s^*=1/2$. }
    \label{fig:glo}
\end{figure}
We first consider evolution under the global NHH $H^{\text{G}}_\text{eff}$. The relevant unnormalised instantaneous right eigenvectors of $H^{\text{G}}_\text{eff}$ are $\ket{\underline{\psi}{}_{\pm} }= -(\delta\mp\Omega)/2g\ket{10} + \ket{01}$, where $\Omega = \sqrt{4g^2+\delta^2}$. The associated eigenvalues take the following form,
\begin{equation}
\xi_\pm = \frac{\delta  + 2\varepsilon \mp \Omega}{2} -i\frac{\Gamma^+}{4}\pm i \frac{\Gamma^+ \delta }{4\Omega},
\end{equation}where $\Gamma^+=\gamma_1^++\gamma_2^-$. We assume $\gamma_1^+ = \gamma^-_2=\gamma$ for simplicity and choose the initial state to be $\ket{\Psi^+}$, allowing us to focus on dynamics within the subspace of eigenstates $\ket{\psi_{\pm}}$. Then the first term in the eigenvalues only contributes to a phase, the second cancels due to normalisation, and the third contributes to the growth or decay of eigenstates. We again quantify state conversion with the fidelity $F_\pm(T) = \braket{\Psi^\pm|\rho_{\text{sd}}(T)|\Psi^\pm}$. Although $H^{\text{G}}_\text{eff}$ exhibits no EP, we can obtain close to perfect CSC between the Bell states $\ket{\Psi^\mp}$ when driving $\delta,\gamma$ according to Eq.~\eqref{eq:traj_mod}, cf. the black curve in Fig. \ref{fig:glo}. A two-step application of the slow-driving operator correctly predicts this effect (see the red curve). We have shown only the CCW trajectory for clarity. 

The picture is drastically different when considering Lindblad dynamics. %The steady state of $\mathcal L^{\text{G}}$ is given by $\rho_{\text{sd}}(s) = (g^2,(g^2+\delta^2)\chi,-g\delta\chi,-g\delta\chi, g^2\chi,g^2\chi)^T/(\delta^2 \chi+g^2(1+\chi)^2)$. 
Under slow-driving, up to the first order, the conversion fidelity takes the form
 \begin{align}
     F_{\pm}(T) = \frac{1}{4}\pm x \frac{ \pi  \delta_0  }{2T\gamma' g}.
 \end{align}At the adiabatic level, we find that the fidelity with the target is only $1/4$. By extension, this imposes that the first-order correction, which depends on the orientation of the trajectory, is physically limited to the range $[0,1/4]$ (such that $F_\pm$ cannot fall below zero), for the slow-driving approximation to be valid. We therefore find that under slow driving, no non-trivial CSC (i.e., with $F_{\pm}>1/2$) can occur between the Bell states $\ket{\Psi^\pm}$ under slow Lindblad dynamics up to the first order.

\section{Conclusions}

In this work, we have introduced the first unified framework that can describe chiral state conversion in any non-Hermitian system described by a hybrid-Lindblad equation. Considering commonly used non-Hermitian models, we showed that CSC is captured by slow-driving predictions in all cases, even when the system shows far-from-adiabatic behaviour somewhere along the trajectory. For NHH evolution, we argued that a two-step application of our slow-driving procedure is needed to obtain state conversion. Importantly, we were able to obtain the precise conversion fidelity at the end of the trajectory, as well as a reasonable match in transient dynamics despite noisy dynamics. For hybrid-Lindblad evolution, we found that even a small amount of quantum jumps in the dynamics remove any need for the two-step procedure; a one-step application of the slow-driving procedure correctly produces the dynamics. In the Lindblad case, we obtained analytical formulas for the conversion fidelity, which can explain the origin of chirality, as well as the physical conditions necessary for state conversion.  

Our results further reveal that at least at the level of slow-driving, encircling the EP is not necessary for CSC. This is in line with numerical and experimental observations made recently \cite{Hassan2017,Kumar2021,Khandelwal2024a,Nasari2022,Kumar2025}. To explore this question further, we considered a model with no EPs. We showed that CSC can be observed therein under NHH dynamics, and can be explained with slow-driving considerations. This proves that the conventional understanding of CSC needs an essential modification; not only does CSC not require encircling an EP or being in its vicinity, it does not require an EP at all. However, in the same model and Lindblad dynamics, CSC is not possible under slow driving. Whether this conflict between the two cases is an artefact or a feature is unclear and may be an interesting direction for further research. 

Lastly, our analysis revealed the possible benefits of fast (non-perturbative) dynamics along the trajectory. We showed that under Lindblad evolution, these effects can lead to a better state conversion fidelity along the chosen orientation, which is especially useful for models that naturally show a lower conversion fidelity (such as model B). However, a complete analysis of non-perturbative effects is beyond the scope of this work and requires a dedicated investigation.

\section*{Acknowledgments}
We thank Patrick Potts for discussions and Géraldine Haack for feedback. E.S. acknowledges support from the Knut and Alice Wallenberg Foundation through the Wallenberg Center for Quantum Technology (WACQT) and the Swedish
Research Council under Contract No. 2023-03498. S.K.~acknowledges support from the Swiss National Science Foundation Grant No. P500PT\textunderscore222265.

\bibliography{References}

\clearpage\newpage
\appendix

\onecolumngrid

\section{Unified framework for slow driving}\label{App:HH}

Here, we show how to obtain the first-order non-adiabatic correction for the evolution equation $\dot\rho(s)=T\mathcal S_s\rho(s)$. To this end, common procedures involve the diagonalisation of the involved evolution operator \cite{Rigolin2008,Rigolin2014,Kumar2025}. We follow the ansatz of \cite{Kumar2025} and express the equation of motion in the eigenbasis of $\mathcal{S}_s$. We now choose to work in vectorised notation, $\underline\rho(s)\to\kett{\underline\rho(s)}$ and $\mathcal S_s\to S_s$, where $S_s$ is the evolution superoperator expressed in matrix form. $S_s$ can be diagonalised as ${S}_s = U^{(0)}_s \Lambda_s V^{(0)}_s$. Here, $\Lambda_s$ is a diagonal matrix with $[\Lambda_s]_{ii} = \lambda_i$, and the columns (rows) of $U^{(0)}_s$ ($V^{(0)}_s$) are the right (left) eigenvectors of ${S}_s$. We now define the rotated state $\kett{\tilde{\rho}(s)} = V^{(0)}_s \kett{\underline\rho(s)} = \sum_k a_k(s) \kett{k}$, where the time-dependent coefficients are given by $a_k(s) = \brakett{\sigma_k}{\underline\rho(s)}$. In this rotated frame, $\kett{\tilde{\rho}(s)}$ is subject to the effective superoperator,
\begin{equation}\label{eq:First_order}
\begin{aligned}
\tilde S_s &=  T \left( V^{(0)}_s {S}_s U^{(0)}_s - \frac{1}{T} V^{(0)}_s  \dot{U}^{(0)}_s \right) \\&= T \left(\Lambda_s - \frac{1}{T} V^{(0)}_s  \dot{U}^{(0)}_s \right).
\end{aligned}
\end{equation}
If the total duration of the protocol is large, i.e., $T\gg 1$, we can treat $X_s \equiv \frac{1}{T} V^{(0)}_s  \dot{U}^{(0)}_s$ as a perturbation to the diagonal matrix $\Lambda_s$. Therefore, we can perform the following decomposition, 
\begin{equation}
\Lambda_s-X_s = U^{(1)}_s \Delta V^{(1)}_s,
\end{equation}
where $\Delta (s) = \Lambda_s + \Lambda'_s$ is a diagonal matrix. In particular, standard quantum perturbation theory up to first order yields that $[\Lambda^\prime_s]_{ii} = -[X_s]_{ii}$, and $U^{(1)}_s = \openone + W_s$ and $V^{(1)}_s = \openone - W_s$ where $W_s = \sum_{n}\sum_{m \neq n} [X_s]_{mn}/(\lambda_m-\lambda_n) \ketbraa{m}{n}$. Using this observation, we define a new state $\kett{\hat{\rho}_s} = V^{(1)}_s \kett{\tilde{\rho}_s}$. In this rotated frame, $\kett{\hat{\rho}_s}$ is subject to the effective superoperator
\begin{equation}\label{eq:First_order_trunc}
\hat{S}_s = T \left(\Delta(s) - \frac{1}{T} V^{(1)}_s  \dot{U}^{(1)}_s \right) = T \Delta(s) + \mathcal{O}(T^{-2}).
\end{equation}
Here we restrict ourselves to first order perturbation, i.e., $\mathcal{O}(T^{-1})$, and find that $\hat{S}_s \approx T \Delta(s)$. This yields that
\begin{equation}
\kett{\hat{\rho}(s)} \approx e^{T \int_0^s \ ds' \Delta(s')} \kett{\hat{\rho}(0)}.
\end{equation}
Note that no time ordering is required since $\Delta$ is diagonal. Transforming the problem back into the original frame yields the state under a slow-driving approximation $\kett{\underline\rho(s)}\approx \kett{\underline\rho_{\text{sd}}}$,
\begin{equation}\label{eq:slow_pert}
\begin{aligned}
\kett{\underline \rho_{\text{sd}}} &=U^{(0)}(s)U^{(1)}(s) e^{T \int_0^s \ ds' \Delta(s')} V^{(1)}(0) V^{(0)}(0) \kett{\rho(0)}\equiv \mathds{U}(s,0)\kett{\rho(0)},
\end{aligned}
\end{equation}where $\mathds{U}(s,0)$ is the slow-driving evolution operator.

\subsubsection{Second-order corrections}

We derive second-order corrections for a general superoperator $\mathcal{S}_s$ in the slow-driving regime. Consider the associated rotated superoperator $\tilde S_s = T\left( \Lambda_s-X_s\right)$ introduced in Eq.\eqref{eq:First_order}, which is applied to the rotated state $\kett{\tilde \rho(s)} = V_s^{(0)}\kett{\rho(s)}$, with $X_s = \frac{1}{T}V_s^{(0)}\dot U_s^{(0)}$. We omit the subscript indicating the time-dependence for notational convenience. In the slow-driving regime where $X \ll \Lambda$, we may perform the following decomposition
\begin{equation}\label{eq:diag_first}
    \Lambda-X = U^{(1)}\Delta^{(1)}V^{(1)},
\end{equation}
where, up to second-order perturbation theory, the entries of the diagonal matrix $\Delta^{(1)}$ are given by
\begin{equation}
    [\Delta^{(1)}]_{ii} = [\Lambda]_{ii}-[X]_{ii} +\sum_{j\neq i} \frac{[X]_{ij}[X]_{ji}}{[\Lambda]_{ii}-[\Lambda]_{jj}}.
\end{equation} 
Note that $[\Lambda]_{ii} = \lambda_i$ is the unperturbed eigenvalue of $S_s$. The associated diagonalising matrices reads,
\begin{equation}
    \begin{aligned}
        U^{(1)} &\equiv \openone + W^{(1,1)} + W^{(1,2)}\\
        V^{(1)} &\equiv \openone - W^{(1,1)} + W^{(1,2)}.
    \end{aligned}
\end{equation}Here, the first and second-order corrections, $W^{(1,1)}$ and $W^{(1,2)}$, take the form
\begin{equation}
\begin{gathered}
W^{(1,1)} =\sum_{n} \sum_{m\neq n} \frac{[X]_{mn}}{[\Lambda]_{mm} -[\Lambda]_{nn}}\dketbra{m}{n} \\
W^{(1,2)}=\sum_n \sum_{m\neq n}\bigg[ \sum_{l \neq n} \frac{[X]_{ml} [X]_{ln}}{([\Lambda]_{mm}-[\Lambda]_{nn})([\Lambda]_{ll}-[\Lambda]_{nn})} -\frac{[X]_{mn} [X]_{nn}}{([\Lambda]_{mm}-[\Lambda]_{nn})^2} \bigg] \dketbra{m}{n}
\end{gathered}
\end{equation}
To exploit the decomposition in Eq.\eqref{eq:diag_first}, we define a new state $\kett{\rho^{(1)}(s)} = V^{(1)}\kett{\tilde \rho(s)}$. In this rotated frame, $\kett{\rho^{(1)}(s)}$ is subject to the effective superoperator,
\begin{equation}
    S^{(1)} = T(\Delta^{(1)}-\underbrace{\frac{1}{T}V^{(1)} \dot U^{(1)}}_{\equiv X^{(2)} }) =  T(\Delta^{(1)}- X^{(2)}).
\end{equation}
Again, we treat $X^{(2)}$ as a perturbation to $\Delta^{(1)}$ and perform the following decomposition,
\begin{equation}
    \Delta^{(1)}-X^{(2)} = U^{(2)}\Delta^{(2)}V^{(2)}.
\end{equation}
When considering second-order corrections  $\mathcal{O}(T^{-2})$, it is sufficient to perform an expansion around $\Delta^{(1)}$ up to first-order perturbation theory, since $\Delta^{(1)}\propto \mathcal O(T^{-1})$ already. This yields that $[\Delta^{(2)}]_{ii} = [\Delta^{(1)}]_{ii}-[X^{(2)}]_{ii}$ and
\begin{equation}
    \begin{gathered}
        U^{(2)} \equiv \openone + W^{(2)}, \qquad V^{(2)}  \equiv \openone - W^{(2)} \\
        W^{(2)} = \sum_{n} \sum_{m\neq n} \frac{[X^{(2)}]_{mn}}{[\Delta^{(1)}]_{mm}-[\Delta^{(1)}]_{nn}}\dketbra{m}{n}.
    \end{gathered}
\end{equation}
Following the procedure described above, we can again define a new state $\kett{\rho^{(2)}} = V^{(2)}\kett{\rho^{(1)}}$, which is subject to a new effective superoperator
\begin{equation}
     S^{(2)} = T(\Delta^{(2)}-\underbrace{\frac{1}{T}V^{(2)}U^{(2)}}_{\equiv X^{(3)} }) =  T(\Delta^{(2)}- X^{(3)}) = T  \Delta^{(2)} + \mathcal{O}(T^{-3}).
\end{equation}
Since we restrict ourselves to second-order perturbation,  we can now truncate the perturbative expansion at this step. Using that $\Delta^{(2)}$ is a diagonal matrix we can solve $\kett{\dot \rho^{(2)}(s)} = T\Delta^{(2)} \kett{\rho^{(2)}(s)}$ as follows,
\begin{equation}
    \kett{\rho^{(2)}(s)} = e^{T \int_0^s ds' \Delta^{(2)}(s')}  \kett{\rho^{(0)}}.
\end{equation}
Moving back to the original frame we obtain the evolved state under slow driving,
\begin{equation}
    \kett{\rho^{\text{sd}}(s)} \equiv \mathbb U(s,0) \kett{\rho(0)} = U^{(0)}(s)U^{(1)}(s) U^{(2)}(s) e^{T \int_0^s ds' \Delta^{(2)}(s')}  V^{(2)}(0) V^{(1)}(0)V^{(0)}(0)\kett{\rho(0)}.
\end{equation}

\subsection{Hamiltonian evolution - recovering the adiabatic theorem}
In the case of Hamiltonian evolution, $\mathcal{S}_s = - i \mathcal{H}_s$, where $\mathcal H_s$ is the time-dependent Hamiltonian superoperator. We further have $\tilde{S}_s = -iT(\Lambda_s -\frac{i}{T} X_s)$ and 
\begin{equation}
\Delta_{kk} = \lambda_k - \frac{i}{T} \brakett{\rho_k}{\dot{\rho}_k},
\end{equation}where $\lambda_k$ are real eigenvalues and $\braa{\sigma_i} = \kett{\rho_i}^\dagger$. The first term contributes to the dynamical phase, while the second to the geometric phase. The evolution operator reads
\begin{equation}\label{eq:HU}
\begin{aligned}
&\mathds{U}^{\text{sd}}(s,0) = \sum_k e^{-iT \int_0^s \ ds' \Delta_{kk}(s') } \ketbraa{\rho_k(s)}{\rho_k(0)}\\
&+ \frac{i}{T}\sum_{m \neq k} e^{-iT \int_0^s \ ds' \Delta_{kk}(s') } \frac{\brakett{\rho_m(s)}{\dot{\rho}_k(s)}}{\lambda_m(s)-\lambda_k(s)} \ketbraa{\rho_m(s)}{\rho_k(0)}- \frac{i}{T}\sum_{m \neq k} e^{-iT \int_0^s \ ds' \Delta_{kk}(s') } \frac{\brakett{\rho_k(0)}{\dot{\rho}_m(0)}}{\lambda_k(0)-\lambda_m(0)}\ketbraa{\rho_k(s)}{\rho_m(0)}.
\end{aligned}
\end{equation}
This is the standard adiabatic approximation along with a first-order non-adiabatic correction. It is equivalent to the adiabatic perturbation theory developed in \cite{Rigolin2008}, generalised for mixed states.

\subsection{Hybrid-Lindblad evolution}

In this case, $\mathcal S_s=\mathcal L_q(s)$. Since $\Delta^q$ is diagonal, we have
\begin{equation}\label{eq:1}
e^{-iT \int_0^s ds' \, \Delta^q(s')} =  \sum_k e^{-iT \int_0^s ds' \, \Delta^q_{kk}(s')} \ketbraa{k}{k}.
\end{equation}Moreover, 
\begin{align}\label{eq:2}
   U^{(0)}(s)= \sum_k \dketbra{\rho^q_k(s)}{k}, \, V^{(0)}(0) = \sum_k \ketbraa{k}{\sigma_k^q(0)}, \,U^{(1)}_s = \openone + W_s,\, \text{and} \,V^{(1)}_s = \openone - W_s,
\end{align} with 
\begin{align}\label{eq:3}
    W_s = \sum_{n}\sum_{m \neq n} \frac{[X]_{mn}}{(\lambda_m-\lambda_n) }\ketbraa{m}{n},
\end{align}where $X_{mk} = \frac{1}{T} \brakett{\sigma_m}{\dot \rho_k}$. Using \eqref{eq:1}, \eqref{eq:2} and \eqref{eq:3} in \eqref{eq:slow_pert} and simplifying, we obtain the evolution operator,
   \begin{equation}
\begin{aligned}
\mathds{U}_q(s,0) =& \sum_k e^{T \int_0^s \ ds' \Delta^q_{kk} (s')} \ketbraa{\rho^q_k(s)}{\sigma^q_k(0)}+ \frac{1}{T}\sum_{m \neq k} e^{T \int_0^s \ ds' \Delta^q_{kk}(s') } \frac{\brakett{\sigma^q_m(s)}{\dot{\rho}^q_k(s)}}{\lambda^q_m(s)-\lambda^q_k(s)} \ketbraa{\rho^q_m(s)}{\sigma^q_k(0)}\\& - \frac{1}{T}\sum_{m \neq k} e^{T \int_0^s \ ds' \Delta^q_{kk}(s') } \frac{\brakett{\sigma^q_k(0)}{\dot{\rho}^q_m(0)}}{\lambda^q_k(0)-\lambda^q_m(0)}\ketbraa{\rho^q_k(s)}{\sigma^q_m(0)}.
\end{aligned}
\end{equation}
Slow driving corresponds to the regime in which non-adiabatic transitions are small. This condition is satisfied when the matrix elements of $\partial_s L_q(s)$ are small compared to the associated energy gap squared, i.e.,
\begin{equation}\label{eq:valid_approx1}
\brakett{\sigma^q_m(s)}{\dot{\rho}^q_k(s)} =  \frac{\brakett{\sigma^q_m(s)| \dot
{{L}}_q(s)}{{\rho}^q_k(s)}}{(\lambda^q_m(s)-\lambda^q_k(s))}
\end{equation}
\begin{equation}\label{eq:valid_approx}
\frac{1}{T}\left| \frac{\brakett{\sigma^q_m(s)}{\dot{\rho}^q_k(s)}}{\lambda^q_m(s)-\lambda^q_k(s)} \right| =\frac{1}{T} \left| \frac{\brakett{\sigma^q_m(s)| \dot
{{L}}_q(s)}{{\rho}^q_k(s)}}{(\lambda^q_m(s)-\lambda^q_k(s))^2} \right| \ll 1 \qquad \text{for all } m\neq k.
\end{equation}

\subsection{Lindblad evolution}
Consider the case of Lindblad evolution $\mathcal S_s= \mathcal L(s)$. We assume that at every point along the parameter trajectory, the Liouvillian has unique steady state and adopt the convention $\lambda_0 = 0$ so that $\dket{\rho_0} = \dket{\rho_{\text{ss}}}$ and $\dbra{\sigma_0} = \dbra{\openone}$ \cite{Minganti2018}. Note that in this case, the exponent $e^{ T \int_0^1 ds' \Delta(s')} \simeq \dketbra{0}{0}$ for large $T$. This yields that
\begin{equation}
U^{(1)} (s) e^{ T \int_0^s ds' \Delta(s')} V^{(1)}(0) \simeq (\openone + W(s))\dketbra{0}{0}(\openone - W(0)) = \left(\dket{0}+\sum_{m \neq 0} \frac{X_{m0}}{\lambda_m}\dket{m}\right)\bigg|_{s' = s} \left(\dbra{0} - \sum_{l\neq 0} \frac{X_{0l}}{\lambda_l}\dbra{l}\right)\bigg|_{s' = 0}.
\end{equation}
Regarding $X_{0l}$, note that
\begin{equation}\label{eq:X_0l}
X_{0l} = -\frac{1}{T}\dbraket{\sigma_0}{\dot{\rho}_l} = -\frac{1}{T}\dbraket{\openone}{\dot{\rho}_l} = -\frac{1}{T}\tr(\dot{\rho}_l) = 0.
\end{equation}
Hence, the expression simplifies to
\begin{equation}
U^{(1)} (s) e^{ T \int_0^s ds' \Delta(s')} V^{(1)}(0) \simeq\dketbra{0}{0}+ \frac{1}{T}\sum_{m \neq 0}  \frac{\dbraket{\sigma_m(s)}{\dot \rho_0(s)}}{\lambda_m(s)} \dketbra{m}{0}
\end{equation}Using the above in Eq. \eqref{eq:slow_pert}, we obtain
\begin{equation}
\begin{aligned}
\mathbb{U}(s,0) =& \left( \sum_k \dketbra{\rho_k(s)}{k} \right) \left(\dketbra{0}{0}+ \frac{1}{T}\sum_{m \neq 0}  \frac{\dbraket{\sigma_m(s)}{\dot \rho_0(s)}}{\lambda_m(s)} \dketbra{m}{0} \right) \left(\sum_l \dketbra{l}{\sigma_l(0)} \right)\\ =& \dketbra{\rho_0(s)}{\sigma_0(0)} + \frac{1}{T}\sum_{m \neq 0} \frac{\dbraket{\sigma_m(s)}{\dot{\rho}_{0}(s)}}{\lambda_m(s)} \dketbra{\rho_m(s)}{\sigma_0(0)}.
\end{aligned}
\end{equation}
Noting that $\dbra{\rho_0}  = \dbra{\rho_{ss}}$ and $\dbra{\sigma_0} = \dbra{\openone}$,  we obtain the final form of the evolution operator,
\begin{equation}
\mathbb{U}(s,0) = \dketbra{\rho_{ss}(s)}{\openone} + \frac{1}{T}\sum_{m \neq 0} \frac{\dbraket{\sigma_m(s)}{\dot{\rho}_{ss}(s)}}{\lambda_m(s)} \dketbra{\rho_m(s)}{\openone}.
\end{equation}
Note that the Drazin inverse is defined as ${L}^+(s) = \sum_{m \neq 0} \frac{\dketbra{\rho_m(s)}{\sigma_m(s)}{}}{\lambda_m(s)}$. Hence, the evolution operator can be written as
\begin{equation}
\mathbb{U}(s,0) = \dketbra{\rho_{ss}}{\openone} +  \frac{1}{T} {L}^+\dketbra{\dot \rho_{ss}}{\openone}.
\end{equation}This first-order correction corresponds precisely to the one obtained in Ref. \cite{Cavina2017}, which suggests that higher orders should be the same. For completeness, we show the consistency of next-order corrections obtained with our approach below.

\subsubsection{Second-order corrections}
Next, we consider second-order corrections to the slow-driving Lindblad evolution operator. Using that $e^{T \int_0^s ds' \Delta^{(2)}(s')} \simeq \dketbra{0}{0}$ for large $T$, we obtain that the evolved state in the original frame,
\begin{equation}
    \kett{\rho^{\text{sd}}(s)} =  \mathbb U(s,0) \kett{\rho(0)} \simeq U^{(0)}(s)U^{(1)}(s) U^{(2)}(s) \dketbra{0}{0} V^{(2)}(0) V^{(1)}(0)V^{(0)}(0)\kett{\rho(0)}.
\end{equation}To simplify the above expression, we start by considering the right expression $\braa{0} V^{(2)} V^{(1)}V^{(0)}$. First, from Eq.~\eqref{eq:X_0l} we have that $[X]_{0n}=\braa{0}V^{(0)}\dot U^{(0)}\kett{n}  = \dbraket{\openone}{\dot \rho_n} = 0$. Using this, we find that
\begin{equation}\label{eq:LF0}
\begin{aligned}
        \braa{0}V^{(1)}&=\braa{0}(\openone+W^{(1,1)}+W^{(1,2)}) \\
        &= \braa{0}-\sum_{n\neq 0} \frac{[X]_{0n}}{[\Lambda]_{nn}}\braa{n} -\sum_{n\neq 0} \bigg[ \sum_{l \neq n} \frac{[X]_{0l} [X]_{ln}}{[\Lambda]_{nn}([\Lambda]_{ll}-[\Lambda]_{nn})} +\frac{[X]_{0n} [X]_{nn}}{[\Lambda]_{nn}^2} \bigg] \braa{n} = \braa{0}.
\end{aligned}
\end{equation}
Second, it follows that
\begin{equation}\label{eq:LS0}
\begin{aligned}
        \braa{0}V^{(2)}&=  \braa{0}(\openone-W^{(2)}) = \braa{0}+\frac{1}{T}\sum_{n\neq 0} \frac{[X^{(2)}]_{0m}}{[\Delta^{(1)}]_{nn}}\braa{n} = \braa{0}+\frac{1}{T}\sum_{n\neq 0} \frac{\braa{0}\dot U^{(1)}\kett{n}}{[\Delta^{(1)}]_{nn}}\braa{n} \\
        &=\braa{0}-\frac{1}{T}\sum_{n\neq 0} \frac{1}{[\Delta^{(1)}]_{nn}}  \partial_s \left( \frac{X_{0n}}{[\Lambda]_{nn}}\right) \braa{n}  + \mathcal{O}(T^{-3}) = \braa{0}.
\end{aligned}
\end{equation}
Here, we have used that $[\Delta^{(1)}]_{00} = 0$. Together, Eq. \eqref{eq:LF0} and \eqref{eq:LS0} imply that 
\begin{equation}
    \braa{0} V^{(2)}(0) V^{(1)}(0)V^{(0)}(0) = \braa{0}V^{(0)}(0) = \braa{\sigma_0(0)} = \braa{\openone}.
\end{equation}
Next, we consider the left expression $U^{(0)}(s)U^{(1)}(s) U^{(2)}(s)\kett{0}$. First, we note that up to the second-order,
\begin{equation}
   U^{(0)}(s) U^{(1)}(s) U^{(2)}(s) \approx U^{(0)}(s)\left(\openone + W^{(1,1)} + W^{(1,2)} + W^{(2)}\right),
\end{equation}where
\begin{align}
W^{(1,1)} \kett{0} &=  \sum_{m\neq 0} \frac{[X]_{m0}}{[\Lambda]_{mm}}\kett{m} = \frac{1}{T} \sum_{m\neq 0} \frac{\dbraket{\sigma_m}{\dot \rho_{0}}}{[\Lambda]_{mm}}\kett{m}\\
 W^{(1,2)} \kett{0} &= \sum_{m\neq 0} \sum_{l \neq 0} \frac{[X]_{ml} [X]_{l0}}{[\Lambda]_{mm}[\Lambda]_{ll}} \kett{m} -\sum_{m\neq0}\frac{[X]_{m0} [X]_{00}}{[\Lambda]_{mm}^2} \kett{m} = \frac{1}{T^2} \sum_{m\neq 0} \sum_{l \neq 0} \frac{\dbraket{\sigma_m}{\dot \rho_l} \dbraket{\sigma_l}{\dot \rho_0} }{[\Lambda]_{mm}[\Lambda]_{ll}} \kett{m}.
\end{align}
Lastly, one obtains that up to second-order perturbation theory,
\begin{equation}
\begin{aligned}
       W^{(2)}\kett{0} &= \frac{1}{T^2} \sum_{m \neq 0}\frac{1}{{[\Delta^{(1)}]_{mm}}}\partial_s \left(\frac{\dbraket{\sigma_m}{\dot \rho_{0}}}{[\Lambda]_{mm}}\right) \kett{m} + \mathcal O (T^{-3})\approx \frac{1}{T^2} \sum_{m \neq 0}\frac{1}{{[\Lambda]_{mm}}}\partial_s \left(\frac{\dbraket{\sigma_m}{\dot \rho_{0}}}{[\Lambda]_{mm}}\right) \kett{m}
\end{aligned}
\end{equation}
Identifying $\kett{\rho_0} = \kett{\rho_{ss}}$, the above leads to
\begin{equation}
    U^{(0)}U^{(1)}U^{(2)}\kett{0} \approx \kett{\rho_{ss}} + \frac{1}{T} \sum_{m\neq 0} \frac{\dbraket{\sigma_m}{\dot \rho_{ss}}}{[\Lambda]_{mm}}\kett{\rho_m} + \frac{1}{T^2} \sum_{m \neq 0} \frac{1}{[\Lambda]_{mm}}\left( \partial_s \left(\frac{\dbraket{\sigma_m}{\dot \rho_{ss}}}{[\Lambda]_{mm}}\right)+\sum_{l \neq 0}\frac{\dbraket{\sigma_m}{\dot \rho_l} \dbraket{\sigma_l}{\dot \rho_{ss}} }{[\Lambda]_{ll}}  \right)\kett{\rho_m}
\end{equation}
Lastly, we note that the second-order correction can be written in terms of the Drazin inverse ${L}^+$,
\begin{equation}
    \frac{1}{T^2} \sum_{m \neq 0} \frac{\dketbra{\rho_m}{\sigma_m}}{[\Lambda]_{mm}}\sum_{l \neq 0} \left( \kett{\rho_l} \partial_s \left(\frac{\dbraket{\sigma_l}{\dot \rho_{ss}}}{[\Lambda]_{ll}}\right) + \frac{\kett{\dot \rho_l} \dbraket{\sigma_l}{\dot \rho_{ss}} }{[\Lambda]_{ll}} \right) = \frac{1}{T^2} {L}^+ \partial_s  \sum_{l \neq 0} \left(\frac{\dketbra{\rho_l}{\sigma_l}}{[\Lambda]_{ll}}\right)\kett{\dot \rho_{ss}} = \frac{1}{T^2} {L}^+ \partial_s {L}^+ \kett{\dot \rho_{ss}}
\end{equation}
Therefore, the second-order evolution operator can expressed as
\begin{equation}
    \mathbb U(s,0) =\dketbra{\rho_{ss}}{\openone} + \frac{1}{T} {L}^+ \dketbra{\dot \rho_{ss}}{\openone} +\frac{1}{T^2} {L}^+ \partial_s {L}^+ \dketbra{\dot \rho_{ss}}{\openone}.
\end{equation}

\subsubsection{Third-order corrections}
Following the procedure outlined in the previous section one may also compute the third-order correction. Doing so, we find that the contribution at the third order reads
\begin{equation}
\begin{aligned}
        &\frac{1}{T^3}\sum_{m \neq 0} \frac{1}{[\Lambda]_{mm}} \partial_s\left( \frac{1}{[\Lambda]_{mm}}\partial_s \left(\frac{\dbraket{\sigma_m}{\dot \rho_{ss}}}{[\Lambda]_{mm}}\right) \right)\dketbra{\rho_m}{\openone} + \frac{1}{T^3}\sum_{m \neq 0} \sum_{k \neq 0}\frac{\dbraket{\sigma_k}{\dot \rho_m}}{[\Lambda]_{kk}[\Lambda]_{mm}} \partial_s \left(\frac{\dbraket{\sigma_m}{\dot \rho_{ss}}}{[\Lambda]_{mm}}\right)\dketbra{\rho_k}{\openone} \\ 
        &+ \frac{1}{T^3}\sum_{m \neq 0} \sum_{k \neq 0}\frac{1}{[\Lambda]_{kk}} \partial_s \left(\frac{\dbraket{\sigma_k}{\dot \rho_m}\dbraket{\sigma_m}{\dot \rho_{ss}}}{[\Lambda]_{kk}[\Lambda]_{mm}}\right)\dketbra{\rho_k}{\openone} + \frac{1}{T^3}\sum_{m \neq 0} \sum_{k \neq 0}\sum_{j \neq 0}\frac{\dbraket{\sigma_j}{\dot \rho_k}\dbraket{\sigma_k}{\dot \rho_m}\dbraket{\sigma_m}{\dot \rho_{ss}}}{[\Lambda]_{jj}[\Lambda]_{kk}[\Lambda]_{mm}} \dketbra{\rho_j}{\openone}
        \end{aligned}
\end{equation}
Exploiting the Drazin inverse, the third-order correction can be simplified to
\begin{equation}
    \frac{1}{T^3}{L}^+\partial_s  L^+ \partial_s {L}^+  \dketbra{\dot \rho_{ss}}{\openone}.
\end{equation}

\section{Chiral state conversion: NHH evolution}\label{App:CSC_NHH}
\subsection{Model A}
\label{App:CSC_NHH1}
\subsubsection{Spectrum of the NHH}
The NHH corresponding to model A is given by
\begin{equation}
H_{\text{eff}}^{\text{A}} = 
\begin{pmatrix}
    \delta -\frac{i}{2}\gamma^- & \kappa \\  \kappa & -\frac{i}{2} \gamma^+
\end{pmatrix}.
\end{equation}
To see the origin of chiral state conversion, we express the eigenvalues in the following form,
\begin{equation}\label{eq:eig_Heff}
\begin{aligned}
    \xi_\pm = \frac{1}{4}\left(2\delta - i\Gamma^+\pm\sqrt{\omega-4i\delta\Gamma^-}\right).
\end{aligned}
\end{equation}
where $\Gamma^\pm = \gamma^-\pm \gamma^+$ and $\omega = 16\kappa^2+4\delta^2-(\Gamma^{-})^2$. We further use that the square root of $z \equiv \omega - 4i\delta \Gamma^-$ can be decomposed into its real and imaginary part,
\begin{equation}
\sqrt{z} = \sqrt{\frac{|z|+\omega}{2}}- i \text{sgn}( \delta \Gamma^-) \sqrt{\frac{|z|-\omega}{2}} = \underbrace{\sqrt{\frac{|z|+\omega}{2}}}_{=\zeta}- i 2 \delta \Gamma^- \underbrace{\sqrt{\frac{2}{|z|+\omega}}}_{=1/\zeta}.
\end{equation}
Then the instantaneous eigenvalues and their associated right eigenstates can be expressed in the form,
\begin{equation}
\begin{aligned}
    &\xi_\pm = \frac{1}{4}(2\delta \pm \zeta) - \frac{i}{4}(\Gamma^+ \pm  \frac{2 \delta \Gamma^-}{\zeta}) \\
    & \ket{\psi_\pm} = \frac{1}{\sqrt{|\Theta_\pm(s)|^2+1}}\left(\ket{0}+\Theta_\pm(s) \ket{1}\right),
\end{aligned}
\end{equation}
where we have defined $\Theta_{\pm} = (i \gamma^-/2+\xi_\pm)/\kappa$. Since the NHH is symmetric, the left eigenvectors are $\bra{\phi_\pm} = \ket{\psi_\pm}^T$.

\subsubsection{Spectrum of the associated superoperator}
Now we consider the spectrum of the associated superoperator $\mathcal L_0^A$, defined through
\begin{equation}
     \mathcal L_0^{\text{A}} \rho = - i  ( H_{\text{eff}}^{\text{A}} \rho - \rho H_{\text{eff}}^{\text{A} \ \dagger}).
\end{equation}The set of eigenvalues, $\{\lambda^0_{lm}\}_{lm}$, and right eigenmatrices, $\{\rho^0_{lm}\}_{lm}$, of $\mathcal L_0^{\text{A}}$ are related to those of $H_{\text{eff}}^{\text{A}}$ through $\lambda^0_{lm} = -i(\xi_m -\xi_l^*)$ and $\rho^0_{lm} = \ketbra{\psi_l}{\psi_m}$ where $l,m \in \{+,-\}$ \cite{Minganti2019}. We therefore find, 
\begin{equation}
\begin{aligned}
&\lambda^0_{\mp \mp} =  -\frac{\Gamma^+}{2}  \pm \frac{\Gamma^- \delta}{\zeta}, \\
&\lambda^0_{\mp \pm} =   -\frac{\Gamma^+}{2}  \pm 2i\zeta,\\ 
&\kett{\rho^0_{\mp \mp}} = \frac{1}{1+|\Theta_\mp|^2}( 1, \Theta_\mp^*, \Theta_\mp, |\Theta_\mp|^2 )^T\\
&\kett{\rho^0_{\mp \pm}} = \frac{1}{\sqrt{(1+|\Theta_+|^2)(1+|\Theta_-|^2)}} ( 1, \Theta_\pm^* ,\Theta_\mp , \Theta_\mp \Theta_\pm^*)^T.
\end{aligned}
\end{equation}

\subsubsection{Chiral state conversion}
Note that $\rho^0_{\mp \pm}$ are non-Hermitian and therefore are not valid quantum states. On the other hand, eigenstates $\rho^0_{\mp \mp}$ are valid quantum states. Since state conversion is not possible for $\gamma^+=\gamma^-$ (i.e., the case of uniform dissipation), we assume $0\leq\gamma^+<\gamma^-$. In this regime, we can evaluate the normalised time-evolved state predicted by the adiabatic approximation,
\begin{equation}
\begin{aligned}
\kett{\rho(s)} &=  \frac{\sum_{l, m }  c_{lm}(0) e^{T\int_0^s ds' \lambda^0_{lm}(s') - \frac{1}{T}\dbraket{\sigma^0_{lm}(s')}{\dot \rho^0_{lm}(s')}}\kett{\rho^0_{lm}(s)}}{\| \sum_{l,m }  c_{lm}(0) e^{T\int_0^s ds' \lambda^0_{lm}(s') - \frac{1}{T}\dbraket{\sigma^0_{lm}(s')}{\dot \rho^0_{lm}(s')}}\kett{\rho^0_{lm}(s)}\|_2} \\
& =  \frac{\sum_{l, m }  c_{lm}(0) e^{T\int_0^s ds' \tilde \lambda^0_{lm}(s') - \frac{1}{T}\dbraket{\sigma^0_{lm}(s')}{\dot \rho^0_{lm}(s')}}\kett{\rho^0_{lm}(s)}}{\| \sum_{l,m }  c_{lm}(0) e^{T\int_0^s ds' \tilde \lambda^0_{lm}(s') - \frac{1}{T}\dbraket{\sigma^0_{lm}(s')}{\dot \rho^0_{lm}(s')}}\kett{\rho^0_{lm}(s)}\|_2},
\end{aligned}
\end{equation}where $\braa{\sigma^0_{lm}}$ denotes the left eigenstates and the effective eigenvalues are
\begin{equation}
    \tilde \lambda^0_{\mp \mp} = \pm \frac{\Gamma^-\delta}{\zeta}, \qquad  \tilde \lambda^0_{\mp \pm} = \pm 2i \zeta.
\end{equation}We have simply factored out the term $e^{-T\int_0^s ds' \frac{\Gamma^+}{2}}$ from the eigenvalues in the numerator and denominator. Note that $\tilde \lambda^0_{\mp \mp}$ are real, whereas $\tilde \lambda^0_{\mp \pm}$ are purely imaginary. This implies that the former will induce exponential growth/decay whereas the latter will only contribute a phase to the normalised state. This is consistent, as the eigenstates corresponding to the latter are not valid quantum states. 

\begin{figure}
    \centering
    \includegraphics[width=0.75\textwidth]{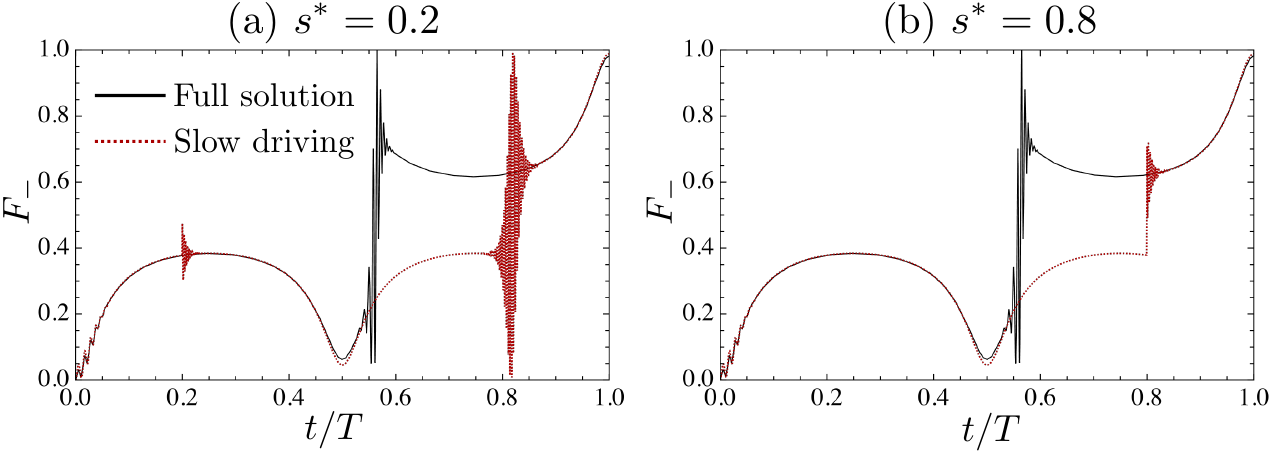}
    \caption{$F_-$ as a function of $s=t/T$ with (a) $s^*=0.2$ and $s^*=0.8$. Parameters taken from Fig. \ref{Fig_1qub_Heff} in the main text.}
    \label{fig:srole}
\end{figure}

In this work we have taken the following periodic driving of the parameters, $\delta(s) = \delta_0 \sin(2\pi x s),\, \gamma(s) = \gamma' + \gamma_0 \sin^2(\pi s)$,
where $\delta_0,\gamma',\gamma_0$ are non-negative constants and $x = \pm 1$ determines the orientation of the trajectory. For a CW trajectory ($x=1)$, $\tilde \lambda^0_{--} (\tilde \lambda^0_{++})$ yields exponential growth (decay) for $s \in [0,1/2]$ and an exponential decay (growth) for $s \in [1/2,1]$, and vice versa for a CCW trajectory $(x=-1)$. Therefore, since $\tilde \lambda^0_{\mp \mp}$ is odd-symmetric around $s = 1/2$, the integral over these effective eigenvalues vanishes at the end point of the trajectory, meaning that
\begin{equation}
    \int_0^1ds' \tilde \lambda^0_{\mp \mp}(s) = 0.
\end{equation}
This implies that the adiabatic theorem fails to predict state conversion for the chosen trajectory. The reason why it fails is that the slow-driving framework, in contrast to exact description with the time-ordered integral, does not capture the ``resetting" of the state at every time step. Taking into account the above observations, the most natural way to solve this problem is to split the application of the slow-driving operator into two time intervals. The evolution of the system from $s=0$ to some $s\leq s^*$ is governed by $\kett{\underline\rho(s)} = \mathbb U^{\text{A}}(s,0)\kett{\rho(0)}$. At $s=s^*$, the system is renormalised and the evolution beyond this point is governed by the operator $\mathbb U^{\text{A}}_0(s,s^*)$. The growth parameters that determine the dominant state in the second half of the trajectory are then
\begin{align}
      \tilde \Delta^0_{\text{\scriptsize{$\pm\pm$}}}(s) = \int_{s^*}^s \, ds' \lambda^0_{\text{\scriptsize{$\pm\pm$}}}(s').
 \end{align}
In Fig. \ref{fig:slowdrivA} in the main text, we chose $s^*=1/2$ and not only found an excellent match between the full solution and the slow-driving prediction in the smooth regions, but also an overlap of the noisy region. This is not expected in all situations for arbitrary choices of $s^*$. We note that while there is considerable freedom in the choice of $s^*$ to obtain correct predictions at the end of the trajectory, the overlap of the noisy region is highly dependent on this choice. In Fig. \ref{fig:srole}, we show the corresponding calculation for $s^*=0.2$ and $s^*=0.8$ with a CCW trajectory, along with the full solution for reference, finding a shift in the noisy region with respect to the full solution. However, these choices still lead to the correct prediction of the dynamics towards the end. 

\subsubsection{Role of the EP}

\begin{figure}
    \centering
    \includegraphics[width=0.75\textwidth]{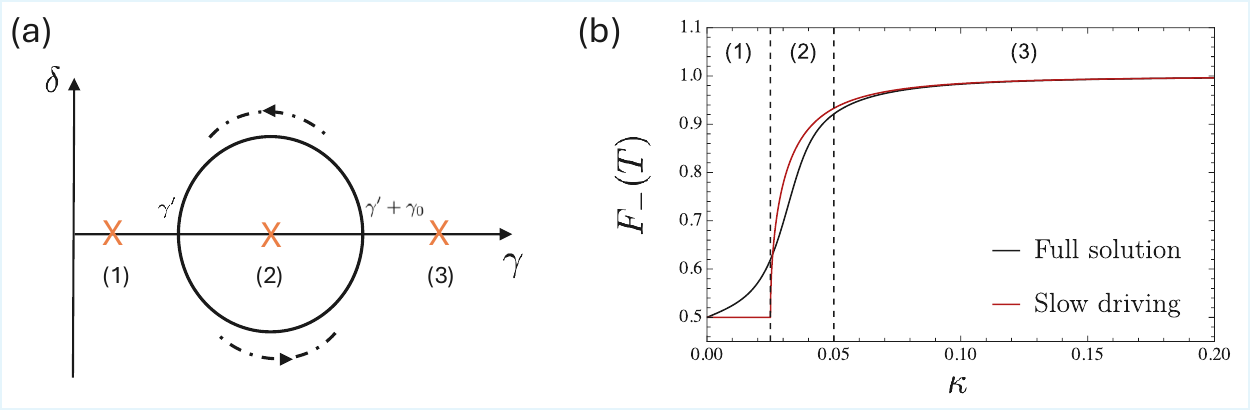}
    \caption{(a) Counterclockwise trajectory in the space of $\delta$ and $\gamma$, and depictions of EPs for different values of $\kappa$ corresponding to the case in which the EP (1) is  on the left of the trajectory, (2) is encircled, (3) is on the right of the trajectory. (b) $F_-$ as a function of $\kappa$. Parameters: $\gamma' = 0.1\delta_0, \gamma_0 = 0.1 \delta_0,\delta_0 = 1, x = -1, T = 20000$.}
    \label{fig:EProle}
\end{figure}

Here, we comment on whether encircling the EP in the NHH is necessary for CSC under slow driving. 
Recall from the main text that the EP of the NHH lies at $\{\delta=0, \kappa = \pm\Gamma^-/4\}$. We consider the CCW trajectory introduced in Eq. \eqref{eq:traj_mod}. For simplicity, we choose $\gamma^+=0$ and $\gamma^-=\gamma(t)$; a similar analysis can be performed for the general case. We take the initial state $\ket{\psi(0)} = \ket{+}$ and the target state $\ket{-}$. To quantify CSC we consider the conversion fidelity at the end point of the trajectory $F_-(T) = |\braket{-|\psi(T)}|^2$, where $\ket{\psi(t)}$ denotes the time-evolved state under NHH evolution. We consider three distinct cases, shown in Fig. \ref{fig:EProle} (a):
\begin{enumerate}
  \item{\textit{Non-encircling trajectory with the EP on the left}: $\kappa<\gamma'/4$.  }
\item{\textit{Encircling trajectory}: $\gamma'/4<\kappa<(\gamma'+\gamma_0)/4$. }
    \item{\textit{Non-encircling trajectory with the EP on the right}: $\kappa>(\gamma'+\gamma_0)/4$. }
\end{enumerate}
If $\kappa = 0$, we have no coupling between the eigenstates of the effective Hamiltonian. As explained in the main text, state conversion is impossible in such a case. However, for $\kappa> 0$, the full solution predicts a non-trivial conversion fidelity; see Fig. \ref{fig:EProle}. Specifically, the conversion fidelity increases monotonically with $\kappa$; the case in which the EP lies on the right of the closed parameter trajectory represents the optimal situation for CSC in model A.

We study how the slow-driving solution compares with the full solution. The slow-driving solution predicts that the initial state will evolve towards the eigenstate with the largest growth parameter. At the final point of the trajectory, $t = T$, the unnormalised eigenstates read
\begin{equation}
\ket{\underline{\psi}{}_\pm} = \ket{0} +\frac{1}{4\kappa}\left(i\gamma' \pm  \sqrt{\gamma'{}^2-16 \kappa^2}\right)\ket{1}.
\end{equation}
The conversion fidelity takes the form,
\begin{equation}\label{eq:trans_fid}
F_-(T) = |\langle -|\psi_\pm\rangle|^2 = \frac{1}{2} \mp \frac{4 \kappa\Re\left(\sqrt{16 \kappa^2-\gamma'{}^2}\right)}{16\kappa^2+|i\gamma' + \sqrt{16 \kappa^2-\gamma'{}^2}|^2}.
\end{equation}
When $\kappa \leq \gamma'/4$, the second term $\Re\left(\sqrt{16 \kappa^2-\gamma'{}^2}\right) = 0$, yielding that $F_-(T) = 1/2$. In other words, for a non-encircling trajectory with the EP on the left, the slow-driving solution does not predict any state conversion. It can be checked that in this region, the mismatch in conversion fidelity between the full solution and slow driving is due to non-perturbative dynamics close to the end of the trajectory. In contrast, when $\kappa >\gamma'/4$ the second term in Eq.~\eqref{eq:trans_fid} is non-zero and CSC is predicted under slow driving. In particular, for $\kappa \gg \gamma'$ the second term approaches $\mp \frac{1}{2}$, predicting perfect CSC. For large  $\kappa$, the full solution and the slow-driving solution match perfectly.

In summary, the EP plays a role in state conversion only to the extent that it can determine whether the involved eigenvalues effectively have real or imaginary elements towards the end of the trajectory. It is not necessary to encircle an EP to obtain CSC, and the effect is dynamical, not topological. 

\subsection{Model B}
\label{app:B}
\subsubsection{Spectrum of the NHH}
The NHH $H_{\text{eff}}^{\text{B}}$, in the basis $\{\ket{11}, \ket{10},\ket{01},\ket{00}\}$, is given by
\begin{equation}
H_{\text{eff}}^{\text{B}} = 
\begin{pmatrix}
    &-\frac{i}{2} \gamma_2^- + \delta + 2 \varepsilon & 0 & 0  &  0 \\
   & 0 &\varepsilon & g &  0 \\
   & 0 & g & -\frac{i}{2}(\gamma_2^-+\gamma^+_1) + \delta + \varepsilon & 0 \\
   &0 &  0  & 0 &-\frac{i}{2}\gamma_1^+
\end{pmatrix}.
\end{equation}
The eigenvalues of $H_{\text{eff}}^{\text{B}}$ take the form,
\begin{equation}
\begin{aligned}
     \xi_0 = -i\gamma_1^+/2, \quad \xi_\pm = \frac{1}{4}\bigg(2\delta + 4\varepsilon - i \Gamma^+ \pm \sqrt{\omega - 4i \Gamma^+ \delta}\bigg), \quad  \xi_1 =-i \gamma_2^-/2+ \delta + 2\varepsilon,
\end{aligned}
\end{equation}where $\Gamma^\pm = \gamma_1^+\pm\gamma_2^-$ $ \omega = 16g^2+ 4 \delta^2 - (\Gamma^{+})^2$. In analogy with model A, we can decompose $ \xi_\pm$ into real and imaginary parts. This yields that
\begin{equation}
  \xi_\pm = \frac{1}{4}(2\delta+4\varepsilon\pm \zeta)-\frac{i}{4}(\Gamma^+\pm \frac{2 \Gamma^+ \delta}{ \zeta}),
\end{equation}
where $ \zeta = \sqrt{(| z| + \omega)/2}$, $ z =  \omega - 4i \Gamma^+ \delta$ and $\omega = 16g^2+ 4 \delta^2-(\Gamma^{+})^2$. We define $ \Theta =  (i\Gamma^+/2-(\delta+ \varepsilon)+ \xi_\pm)/g$. The NHH only contains coupling terms between single-excitation states, $\ket{01}$ and $\ket{10}$, whereas the ground state, $\ket{00}$, and the doubly excited state, $\ket{11}$, remain uncoupled. As a consequence of this coupling structure, the  ground and excited state, $\ket{00}$ and $\ket{11}$, are themselves eigenvectors, while the remaining two eigenvectors are superpositions of $\ket{01}$ and $\ket{10}$. Specifically, the normalised eigenstates take the following form
\begin{equation}\label{eq:Eig_2qubit}
\begin{aligned}
\ket{\psi_0}=\ket{00}, \quad \ket{\psi_\pm} = \frac{1}{\sqrt{|\Theta_\pm|^2+1}}\left( \ket{01} +  \Theta_\pm \ket{10}\right), \quad \ket{\psi_{1}}=\ket{11}.
\end{aligned}
\end{equation}

As the NHH does not couple the eigenvectors $\ket{00}$ and $\ket{11}$, their time evolution is trivial up to a phase factor, preventing any state conversion between them. Hence, we may only observe chiral state conversion between $\ket{\psi_\pm}$. This can be seen by studying the full dynamical evolution under $H_{\text{eff}}^{\text{B}}$. We start by discretising the time-ordered exponential into $N$ time-steps of length $\Delta s$. Let the initial state be $\ket{\psi(0)}$. Then at a later time, $s = t/T$ the unnormalised state is given by
\begin{equation}
\ket{ \underline\psi(s)} = \mathcal{T} e^{-i  T \int_0^s H_{\text{eff}}^{\text{B}}(s')ds'}\ket{\psi(0)} \approx  e^{-i T H_{\text{eff}}^{\text{B}}((N-1) \Delta s)\Delta s}\dots  e^{-i T H_{\text{eff}}^{\text{B}}(0)\Delta s}\ket{\psi(0)} = \prod_{k=0}^{ N-1} e^{-i T H_{\text{eff}}^{\text{B}}(k\Delta s)\Delta s}\ket{\psi(0)},
\end{equation}
where the matrix product is time-ordered and $N \Delta s = s$. Away from an EP we can diagonalise the instantaneous NHH at each time-step $k\Delta s$, i.e.,
\begin{equation}
H^{\text{B}}_{\text{eff}}(k \Delta s)= P(k \Delta s) D(k \Delta s) P^{-1}(k \Delta s),
\end{equation}
where the columns of $P$ correspond to the right instantaneous eigenstates, the rows of  $P^{-1}$ correspond to the left instantaneous eigenstates, and $D$ is the diagonalised NHH at time $k\Delta s$. $P$ and $P^{-1}$ take the form
\begin{equation}
P(s) = 
\begin{pmatrix} &1 & 0& 0&0  \\ 
&0  & \frac{\Theta_{+}}{\sqrt{| \Theta_{+}|^2+1}}& \frac{ \Theta_{-}}{\sqrt{| \Theta_{-}|^2+1}}  & 0 \\ 
&0& \frac{1}{\sqrt{| \Theta_{+}|^2+1}}  &\frac{1}{\sqrt{| \Theta_-|^2+1}} & 0 \\
& 0 &0 &0 & 1
\end{pmatrix}, \qquad P^{-1}(s) = 
\begin{pmatrix} &1 & 0& 0&0  \\ 
&0  &  - \frac{g  \sqrt{| \Theta_+|^2+1}}{\xi_- -\xi_+} & \frac{g \Theta_-  \sqrt{| \Theta_+|^2+1}}{\xi_- -\xi_+} & 0 
\\ &0 & \frac{g \sqrt{| \Theta_-|^2+1}}{\xi_- -\xi_+}   & - \frac{g \Theta_+  \sqrt{| \Theta_-|^2+1}}{\xi_- -\xi_+} & 0 \\
& 0 &0 &0 & 1
\end{pmatrix}
\end{equation}
Now, we can express the diagonal decomposition as follows 
\begin{equation}
P(s) = 
\begin{pmatrix} 
&1 & 0&0  \\ 
&0  & \mathcal P(s) & 0 \\
& 0 &0  & 1
\end{pmatrix}, \qquad P^{-1}(s) = 
\begin{pmatrix} &1 & 0& 0\\ 
&0 & \mathcal P^{-1}(s) & 0 \\
& 0 &0  & 1
\end{pmatrix}, \qquad D(s) =  \begin{pmatrix} 
&\xi_1(s) & 0& 0\\ 
&0 & \Xi(s) & 0 \\
& 0 &0  & \xi_0(s)
\end{pmatrix}
\end{equation}
where $ \Xi(s) = \text{diag}(\xi_+,\xi_-)$. Hence, the time-ordered evolution in the eigenbasis reads 
\begin{equation}
\begin{aligned}
\ket{\underline\psi(s)} &= \prod_{k=0}^{N-1} e^{-i H_{\text{eff}}^{\text{B}}(k\Delta s)\Delta s}\ket{\psi(0)}\\ &=  \begin{pmatrix} e^{-i T \sum_{k=0}^{N-1} \xi_1 ( k \Delta s) \Delta s } & 0 & 0 
\\ 0 &\prod_{k=0}^{N-1} \mathcal P( k \Delta s) e^{-i T \ \Xi(k \Delta s)  \Delta s} \mathcal P^{-1}( k \Delta s) & 0\\ 
0 & 0 &  e^{-i T \sum_{k=0}^{N-1} \xi_0 ( k \Delta s) \Delta s } \end{pmatrix} \ket{\psi(0)}
\end{aligned}
\end{equation}

Let us first comment on the role of the initial state. To observe state conversion, the initial state must have a non-zero overlap with the subspace spanned by $\{\ket{01},\ket{10}\}$. If we initialise in $\ket{00}$ or $\ket{11}$, the state will remain in these eigenvectors up to a trivial phase, as these 
states remain decoupled from the rest throughout the evolution. Moreover, the NHH that governs the dynamics in the subspace spanned by $\{\ket{10}, \ket{01}\}$ is given by
\begin{equation}
H_{\text{eff}}^{\text{sub}} = \begin{pmatrix}\varepsilon & g  \\ g & \varepsilon+\delta -\frac{i}{2}\Gamma^+ \end{pmatrix}
\end{equation}
This resembles $H_{\text{eff}}^{\text{A}}$, up to an energy scaling. This implies that the physical mechanisms behind CSC in model A and B will be similar, leading to similar results for the two models.

\subsubsection{Spectrum of the associated superoperator and chiral state conversion}
Next, we study the spectrum of $\mathcal L_0^{\text{B}}$, where the superoperator is defined as
\begin{equation}
\dot \rho = \mathcal L_0^{\text{B}} \rho = -i(H_{\text{eff}}^{\text{B}} \rho- \rho H_{\text{eff}}^{\text{B} \ \dagger }).
\end{equation}To investigate dynamics with the superoperator $\mathcal L_0^{\text{B}}$, we restrict to the steady-state subspace, which can be shown to be the most relevant \cite{Khandelwal2021}. Specifically, the steady-state subspace is spanned by the basis matrices $\{\ketbra{11}{11}, \ketbra{10}{10},\ketbra{10}{01},\ketbra{01}{10}, \ketbra{01}{01},\ketbra{00}{00}\}$, giving rise to a $6\times 6$ superoperator. The eigenvalues take the form
\begin{equation}
\begin{aligned}
        &\lambda^0_{\scriptsize{\text{00}}}(s) = -i({\xi_0}-  {\xi_0}^* ) = -\gamma_1^+ \\
        &\lambda^0_{--}(s) = -i({\xi_-}-  {\xi_-}^* ) = - \frac{\Gamma^+}{2}\left(1-\frac{2 \delta}{ \zeta}\right) \\
        &\lambda^0_{-+}(s) = -i({\xi_-}-  {\xi_+}^* ) = - \frac{\Gamma^+-i \zeta}{2} \\
        &\lambda^0_{+-}(s) = -i({\xi_+}-  {\xi_-}^* ) = - \frac{\Gamma^++i \zeta}{2} \\
        &\lambda^0_{++}(s) = -i({\xi_+}-  {\xi_+}^* ) =- \frac{\Gamma^+}{2}\left(1+\frac{2 \delta}{ \zeta}\right) \\
        &\lambda^0_{11}(s) = -i({\xi_1}-  {\xi_1}^* ) = -\gamma_2^- .
\end{aligned}
\end{equation}
The associated eigenmatrices are
\begin{equation}
    \begin{aligned}
        &\rho^0_{00} = \ketbra{00}{00} \\
        & \rho^0_{--} = \ketbra{\psi_-}{\psi_-} = \frac{1}{|\Theta_-|^2+1}\left( \ketbra{01}{01} +\Theta_-^*\ketbra{01}{10}+ \Theta_-\ketbra{10}{01}+ |\Theta_-|^2\ketbra{10}{10} \right)\\
        & \rho^0_{-+} = \ketbra{\psi_-}{\psi_+} = \frac{1}{\sqrt{(| \Theta_+|^2+1)(|  \Theta_-|^2+1)}}\left( \ketbra{01}{01} + \Theta_+^*\ketbra{01}{10}+\Theta_-\ketbra{10}{01}+ \Theta_-\Theta_+^*\ketbra{10}{10} \right)\\
        & \rho^0_{+-} = \ketbra{\psi_+}{\psi_-} =\frac{1}{\sqrt{(| \Theta_+|^2+1)(| \Theta_-|^2+1)}}\left( \ketbra{01}{01} +\Theta_-^*\ketbra{01}{10}+ \Theta_+\ketbra{10}{01}+ \Theta_-^*\Theta_+\ketbra{10}{10} \right)\\
        & \rho^0_{++} = \ketbra{\psi_+}{\psi_+} = \frac{1}{|\Theta_+|^2+1}\left( \ketbra{01}{01} + \Theta_+^*\ketbra{01}{10}+ \Theta_+\ketbra{10}{01}+ |\Theta_+|^2\ketbra{10}{10} \right)\\
        &\rho^0_{11} = \ketbra{11}{11}. \\
    \end{aligned}
\end{equation}

\begin{figure}
    \centering
    \includegraphics[width=0.75\textwidth]{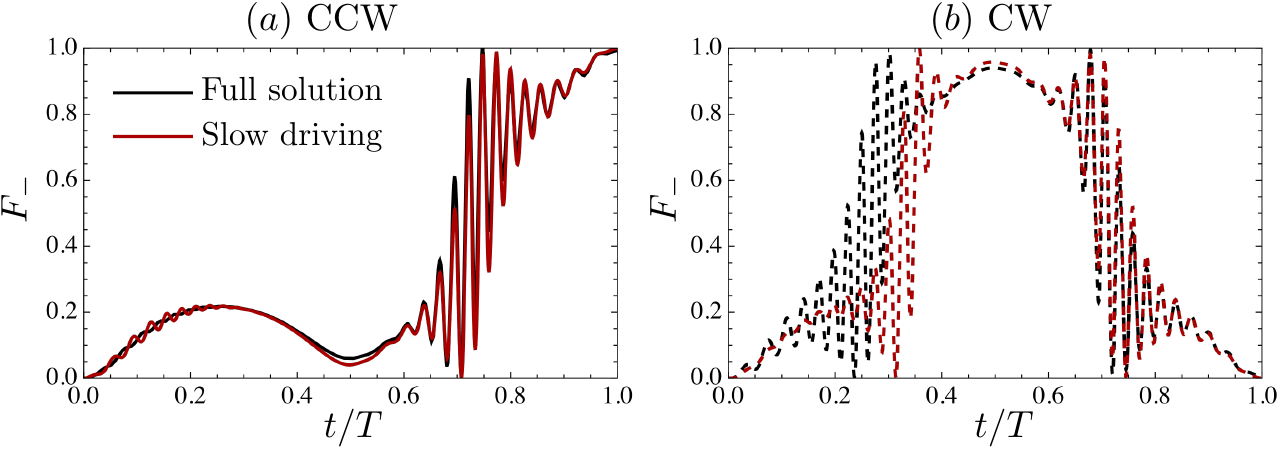}
    \caption{$F_-$ as a function of $s=t/T$ for model B with initial state $\ket{\Psi^+}$ and equal dissipation rates $\gamma^+_1 = \gamma_2^-=\gamma(t)$. Parameters: $\gamma' = 0.025 \delta_0$, $\gamma_0 = 0.25 \delta_0$, $g = 0.35 \delta_0$, $\delta_0 = 0.04$, $\epsilon =1 $ and $\gamma' T = 5$.}
    \label{fig:twoqubit}
\end{figure}

Lastly, we consider the initial state $\ket{\Psi^+}$ and study the conversion fidelity $F_-(s) = \bra{\Psi^-}\rho(s)\ket{\Psi^-}$ as a function of $s = t/T$ for CCW $(x=-1)$ and CW $(x=1)$ trajectories. The result is shown in Fig. \ref{fig:twoqubit}. Here, we compare the exact solution (black) with the slow driving prediction (red), where the latter is obtained by using the two-step application of the slow-driving evolution operator.

\section{Chiral state conversion: Hybrid-Lindblad evolution}
\label{app:hyb}
Here, we comment on the validity of the slow-driving evolution operator $\mathbb{U}_{q=0.5}^{\text{A}}(s,0)$ along the trajectory \eqref{eq:traj_mod}. From Eq.~\eqref{eq:valid_approx} we have that the slow-driving assumption is valid when the matrix elements of $\dot{\mathcal{L}}_q(s)$ are small compared to the associated energy gap squared. However, close to the end of the trajectory, we find that this condition breaks. We denote the eigenvalue associated with the largest growth parameter by $\lambda^q_*$ and its associated right and left eigenstate by $\kett{\rho^q_*}$ and $\braa{\sigma^q_*}$, respectively. We refer to the remaining eigenvalues as $\lambda^q_m$, with associated right and left eigenvectors $\kett{\rho_m}$ and $\braa{\sigma^q_m}$, where $m = 1,2,3$. At the end of the trajectory, the growth parameter $\Delta^q_*$ associated $\lambda^q_*$ is at least one order of magnitude larger than $\Delta^q_m$. Hence, at the end of the trajectory, the slow-driving evolution operator is well approximated by
\begin{equation}
\begin{aligned}
U_{0.5}^{\text{A}}(s,0)\approx\,   e^{T \int_0^s \ ds' \Delta^q_*(s')} \bigg(& \ketbraa{\rho^q_*(s)}{\sigma^q_*(0)}+ \frac{1}{T}\sum_{m = 1}^3 \frac{\brakett{\sigma^q_m(s)}{\dot{\rho}^q_*(s)}}{\lambda^q_m(s)-\lambda^q_*(s)} \ketbraa{\rho^q_m(s)}{\sigma^q_*(0)}\\& - \frac{1}{T}\sum_{m =1}^3  \frac{\brakett{\sigma^q_*(0)}{\dot{\rho}^q_m(0)}}{\lambda^q_*(0)-\lambda^q_m(0)}\ketbraa{\rho^q_*(s)}{\sigma^q_m(0)}\bigg).
\end{aligned}
\end{equation}

Accordingly,, using Eq. \eqref{eq:valid_approx}, slow driving is characterised by the coefficients
%For $\mathbb{U}_{0.5}^{\text{A}}(s,0)$ to satisfy the slow-driving assumption, we require that
\begin{equation}
   M_m \equiv \frac{1}{T} \left| \frac{\brakett{\sigma^q_m(s)}{\dot{\rho}^q_*(s)}}{\lambda^q_m(s)-\lambda^q_*(s)}\right| \ll 1, \qquad \text{for } m = 1,2,3.
\end{equation}

\begin{figure}
    \centering
    \includegraphics[width=0.4\textwidth]{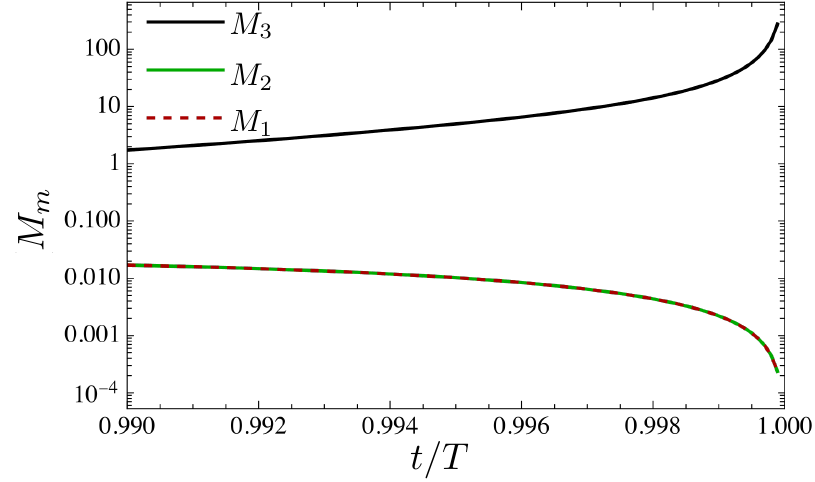}
    \caption{$M_m$ as a function of $s=t/T$. Parameters taken from Fig. \ref{fig:q} in the main text.}
    \label{fig:Mplot}
\end{figure}

In Fig. \ref{fig:Mplot}, we plot the above coefficients as a function of time for $s \in [0.99,1]$ for $m = 1,2,3$. We find a sharp increase in one of the coefficients, $M_3$, as $s\to1$, breaking the slow-driving assumption. We note that this discrepancy only arises very close to the end of the trajectory $s\sim1$ and is highly dependent on its form \eqref{eq:traj_mod}, which we have chosen.  Moreover, this issue does not arise for $q=0,1$, in which case slow-driving correctly predicts the final state at $s=1$.

\section{Chiral state conversion: Lindblad evolution}\label{app:slowdriv}
\begin{figure}
    \centering
    \includegraphics[width=0.4\textwidth]{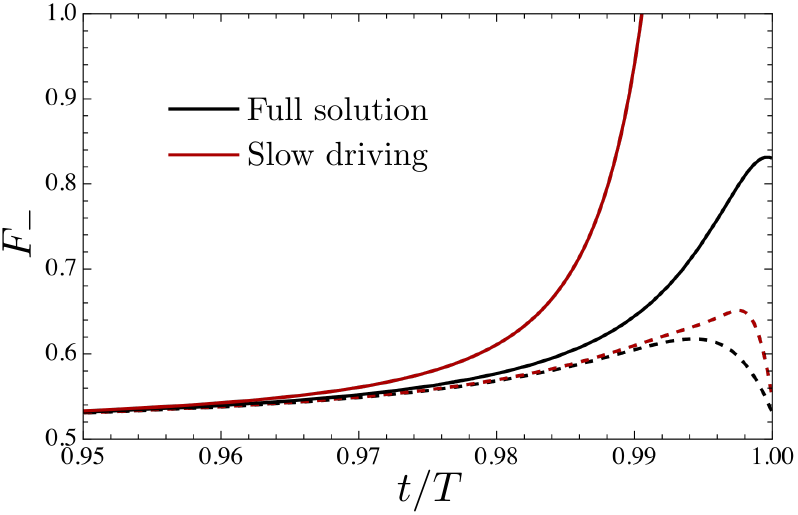}
    \caption{$F_-$ (with $\rho(0)=\ketbra{\Psi^+}{\Psi^+}$ and a CCW trajectory) as a function of $s=t/T$ for model B. We have taken $\gamma_1^+=\gamma_2^-=\gamma(t)$. Parameters: $\delta_0=1$, $g=0.01\delta_0$, $\gamma_0=0.01\delta_0$, $\gamma'=0.025\delta_0$ (dashed), $\gamma'=0.001\delta_0$ (solid) and $T=17000$.}
    \label{fig:lindB}
\end{figure}

In Fig. \ref{fig:lindB}, show $F_-$ (with $\rho(0)=\ketbra{\Psi^+}{\Psi^+}$ and a CCW trajectory) as a function of $s=t/T$ for model B. In both sets of curves, we find an excellent match for intermediate times, so the curves are shown only at the end of the trajectory, where there is a mismatch. In the dashed curves ($\gamma'=0.025\delta_0$), there is a reasonable match between the full solution and the first-order slow-driving prediction. However, the obtained final fidelity is quite small. On the other hand, for the solid curves ($\gamma'=0.001\delta_0$), there is a large mismatch and the dynamics evidently become non-perturbative towards the end of the trajectory. However, here, we obtain a considerably higher conversion fidelity. The reason for this non-perturbative behaviour is that as $\gamma'\to0$, a degeneracy appears in the Liouvillian at $s=1$. This breaks an assumption of the slow-driving framework. Let us also note that the second-order correction gives a worse prediction in both cases.

Below, we give the expressions for the conversion fidelity up to the next order. 
\subsubsection{Model A}
\begin{equation}
\begin{aligned}
&F_{\pm}(T) =\frac{1}{2}\pm x \frac{16 \pi \kappa  \delta _0 }{T\gamma'(8  \kappa^2 +\gamma'^2) }  \\&\mp x\frac{256 \pi ^3 \delta_0 k  \left(\gamma'^5 \left(\gamma'^2+5 \gamma' \gamma_0+60 \delta_0^2\right)+256 k^6 (2 \gamma'+\gamma_0)+16 \gamma' k^4 \left(12 \gamma'^2+11 \gamma' \gamma_0+88 \delta_0^2\right)+8 \gamma'^3 k^2 \left(3 \gamma'^2+5 \gamma' \gamma_0+64 \delta_0^2\right)\right)}{T^3\gamma'^4\left(8  k^2+\gamma'^2 \right)^4}
\end{aligned}
\end{equation}

\subsubsection{Model B}
\begin{align}
    F_{\pm} (T) =  \Big(\frac{1}{2}-\frac{g^2}{4 g^2+\gamma'^2}\Big) \pm x \frac{2 \pi  g\delta_0 }{T\gamma'(4g^2+\gamma'^2)} +\frac{4 \pi ^2 g^2 \left(\gamma'^4 \left(3 \gamma' \gamma_0+11 \delta_0^2\right)+16 \delta_0^2 g^4-4 \gamma'^2 g^2 \left(\gamma' \gamma_0-4 \delta_0^2\right)\right)}{ T^2 \gamma'^2\left(4 g^2+\gamma'^2\right)^4}
\end{align}

\section{Model B - global master equation}\label{app:globalME}

\subsection{Lindblad jump operators}
The eigenvalues of the system Hamiltonian $H^{\text{B}}$ are $\varepsilon_0=0$, $\varepsilon_\pm= \big(\delta+2\varepsilon \pm \sqrt{4g
^2+\delta^2}\big)/2$, and $\varepsilon_1 = 2\varepsilon+\delta $, and the corresponding unnormalised eigenstates are $\ket{00}$, $\ket{\underline{\varepsilon}_\pm} = (-\delta \pm \Omega)/2g\ket{10} + \ket{01}$, and $\ket{11}$, respectively. The non-zero Lindblad operators are the ones corresponding to transitions of energies $\varepsilon_-$ and $\varepsilon_+$. For each of these energies, there are two possible transitions, as $\varepsilon_-=\varepsilon_--\varepsilon_0=\varepsilon_1 - \varepsilon_+$ and $\varepsilon_+=\varepsilon_+-\varepsilon_0=\varepsilon_1-\varepsilon_-$ \cite{Hofer2017}. The jump operators are thus given by \cite{Hofer2017}
\begin{equation}
	\begin{aligned}
		& L_k(\varepsilon_-) = \ket{00}\!\bra{00}\sigma_-^{\left(k\right)}\ket{\varepsilon_-}\!\bra{\varepsilon_-} + \ket{\varepsilon_+}\!\bra{\varepsilon_+}\sigma_-^{\left(k\right)}\ket{11}\!\bra{11}\\
		&L_k(\varepsilon_+) = \ket{00}\!\bra{00}\sigma_-^{\left(k\right)}\ket{\varepsilon_+}\!\bra{\varepsilon_+} + \ket{\varepsilon_-}\!\bra{\varepsilon_-}\sigma_-^{\left(k\right)}\ket{11}\!\bra{11},
	\end{aligned}
\end{equation}
and the adjoints of the above.

\subsection{Spectrum of the effective Hamiltonian}
Here, we show that the two-qubit global NHH $H_{\text{eff}}^{\text{G}}$ has no EP. Its eigenvalues take the following form,
\begin{equation}
\begin{aligned}
&\xi_0  = -\frac{i\gamma_1^+}{2}\\
&\xi_2 = \delta + 2\epsilon - \frac{i\gamma_2^-}{2}\\
&\xi_- =
\frac{\delta  + 2\epsilon+ \Omega}{2}-i \left(\frac{\Gamma^+}{4}+\frac{ \Gamma^+\delta}{4\Omega}\right)\\
&\xi_+ = \frac{\delta  + 2\epsilon- \Omega}{2}-i \left(\frac{\Gamma^+}{4}-\frac{ \Gamma^+\delta}{4\Omega}\right),
\end{aligned}
\end{equation}where $\Gamma^+ = \gamma_1^++\gamma_2^-$ and $\Omega = \sqrt{4g^2+\delta^2}$. It is clear that for nonzero $g,\delta$, the NHH does not have an EP. The only point where the eigenvalues $\xi_{\pm}$ merge is the trivial situation when $g = 0$ and $\delta = 0$. However, in this case the NHH becomes diagonal, which implies that it has a complete set of eigenvectors. Therefore, the global NHH has no EP.

\end{document}